\documentclass[12pt, reqno]{amsart}

\newtheorem{theorem}{Theorem} %[section]
\newtheorem{lemma}{Lemma}

\theoremstyle{definition}
\usepackage{setspace}
\usepackage{nomencl}
\usepackage{tabularx}
\makenomenclature
\usepackage{algorithm2e}
\usepackage[round]{natbib} 
\usepackage{epsfig}
\usepackage{rotating}
\usepackage{url}
\usepackage{mathtools}
\usepackage{amsfonts}
\usepackage{comment}
\usepackage{mathrsfs}
\usepackage{textcomp}
\usepackage{bm}
\usepackage{titletoc}
\usepackage[OT1]{fontenc}
\usepackage{graphicx,epsfig}
\usepackage{epstopdf}
\usepackage{amscd}
\usepackage[caption=false]{subfig}
\usepackage{amsmath,latexsym,amssymb,amsthm, esint}
\usepackage{hyperref}
\hypersetup{colorlinks=true,citecolor=blue,linktocpage=true }
\usepackage[utf8]{inputenc}
\usepackage{mathtools}

\newtheorem{rmk}{\bf Remark}%[section]
\usepackage{multirow}
\newpage

\begin{document}

%\title[Change point analysis of angular data]{Intrinsic statistical framework to detect change point in angular data}
% \title{
% Change point detection in  angular data\\ with the new  measure of variation arising from the intrinsic geometry of curved torus  }

% \title[Exact uniform sampling from curved torus]{An efficient algorithm for rejection-free area uniform sampling from torus using its intrinsic geometry}

\title[Exact uniform sampling from curved torus]{Intrinsic geometry-inspired dependent toroidal distribution: Application to regression models for astigmatism data}
\maketitle
\begin{center}

\author{Buddhananda Banerjee}\\
\email{bbanerjee@maths.iitkgp.ac.in}\\
\address{Department of Mathematics,  Indian Institute of Technology Kharagpur, India-$721302$}\\

and \\

\author{Surojit Biswas}\\ %\footnote{ Advisor: Dr. Buddhananda Banerjee}} \\
\email{surojit23$@$iitkgp.ac.in}\\
\address{Department of Mathematics, Indian Institute of Technology Kharagpur, India-$721302$}

\end{center}

\begin{abstract}

This paper introduces a dependent toroidal distribution, to analyze astigmatism data following cataract surgery. Rather than utilizing the flat torus, we opt to represent the bivariate angular data on the surface of a curved torus, which naturally offers smooth edge identifiability and accommodates a variety of curvatures-- positive, negative, and zero. Beginning with the area-uniform toroidal distribution on this curved surface, we develop a five-parameter-dependent toroidal distribution that harnesses its intrinsic geometry via the area element to model the distribution of two dependent circular random variables. We show that both marginal distributions are Cardioid, with one of the conditional variables also following a Cardioid distribution. This key feature enables us to propose a circular-circular regression model based on conditional expectations derived from circular moments. To address the high rejection rate (approximately 50\%) in existing acceptance-rejection sampling methods for Cardioid distributions, we introduce an exact sampling method based on a probabilistic transformation. Additionally, we generate random samples from the proposed dependent toroidal distribution through suitable conditioning. This bivariate distribution and the regression model are applied to analyze astigmatism data arising in the follow-up of one and three months due to cataract surgery.

 % \textcolor{blue}{ This article presents an intrinsic geometrical approach to address change point problems in angular data for concentration parameters and mean direction separately. A new measure termed ``curved variance," analogous to variance for angular random variables, is introduced. Utilizing this "curved variance," two tests are developed. The first one is to identify the change point in the concentration parameter having pivotal distribution free from concentration and mean direction parameters, which gives better power than the existing one, and the second one is to identify the change point in the mean direction which performs equivalently to existing test for von Mises distribution but with substantially reduced computational complexity. Notably, both tests are extendable to other unimodal circular distributions.}

% \textit{(Diaconis, P., Holmes, S., & Shahshahani, M. (2013). Sampling from a manifold. Advances in modern statistical theory and applications: a Festschrift in honor of Morris L. Eaton, 10, 102-125.)} 

\vspace{0.5cm}

\end{abstract}

\newpage

\tableofcontents{}

\nomenclature{\(\mathbb{R}^n\)}{n-dimensional real space}
\nomenclature{\(\mathbb{C}\)}{Space of complex numbers}
\nomenclature{\(\mathbb{N}\)}{Set of natural numbers}
\nomenclature{\(A^t\)}{Transpose of matrix A}

\nomenclature{\(\mathbb{S}^1\)}{Unit circle}
\nomenclature{\(\Omega\)}{Sample space}
\nomenclature{\(\mathcal{F}\)}{\sigma-Field}
\nomenclature{\(P\)}{Probability measure}
\nomenclature{\(\mathbb{P}\)}{Projection}
\nomenclature{\(\Omega\)}{Sample space}
\nomenclature{\(\mathbb{T}_a^{n}\)}{n-dimensional flat torus}
\nomenclature{\(\mathbb{T}_m^{n}\)}{n-dimensional general torus}
\nomenclature{\(\theta_1\)}{Horizontal angle}
\nomenclature{\(\theta_2\)}{Vertical angle}
\nomenclature{\(\mathcal{S}\)}{Parameter space}

%%%%%%%%%%%%%^ EXTENSION SEMINAR FORM 4TH TO 5TH YEAR %%%%%%%%%%%%%%%%%%%%%%%%%%%%

\tableofcontents{}

\newpage 

%%%Intro
% Data story
% necessity of geometry of toridal data
%%%%%Sec- dist on surface of trous
%Area uniform
%dependent distr
%%%Sec-Estimation
%%%%sec-regression,con,marginal
%%%%Sec-Sampling AND SImulation
%%%Data Analysis

\section{Introduction}

Analyzing data from a manifold requires careful attention to how the data is sampled and represented, as these factors significantly impact the statistical inferences drawn from the data, such as parameter estimation, hypothesis testing, and prediction. This paper specifically examines the 2-dimensional curved torus $(\mathbb{S}^1 \times \mathbb{S}^1)$, a Riemannian manifold embedded in \( \mathbb{R}^3 \). In contrast to circular and spherical data, which are also presented on Riemannian manifolds and naturally incorporate the geometry of the respective surfaces in statistical analysis (as discussed by  \cite{mardia2000directional}), the flat torus,  $(0, 2\pi] \times (0,2\pi]$, does not provide the same. 
The flat torus and the curved torus are not homeomorphic due to differences in their topological properties. This advocates the importance of choosing the correct geometric representation while working with data that inherently resides on a toroidal surface. Hence we prefer to represent the bivariate angular data on the surface of a curved torus, which enjoys smooth identifiability of the edges leading to a continuous regression model rather than many piece-wise continuous functions for the same on a flat torus.

The bivariate von Mises density, introduced by  \cite{mardia1975statistics}, is one of the most well-known bivariate circular distributions on the flat torus. Over time, more parsimonious submodels have been developed by various researchers, including \cite{rivest1988distribution}, \cite{singh2002probabilistic}, \cite{mardia2007protein},  \cite{kent2008modelling}, and \cite{ameijeiras2022sine}. For a broader understanding of toroidal distributions on the flat torus, readers can refer to the work by   \cite{ley2017modern}.
It is crucial to recognize that most of the existing studies have focused on the flat torus. 
It implies that statistical techniques developed for the flat torus may not be applicable to the curved torus. Hence a separate analysis of the data is required on the surface of the curved torus.
While several distributions have been studied on the flat torus, the curved torus, with its more intriguing topological structure in \( \mathbb{R}^3 \), has seen limited exploration. As far as we know, only two studies have specifically addressed distributions on the curved torus: one by  \cite{diaconis2013sampling} on the uniform toroidal distribution, and another by  \cite{biswas2024exploring} on a maximum entropy distribution that extends the von Mises distribution to the curved torus.

In this paper, we build on previous work by developing a dependent toroidal distribution derived by transforming the uniform toroidal distribution on a curved torus. This new distribution considers the topology of the manifold, offering a more natural representation for analyzing toroidal data. Our investigation is structured across several sections. Section \ref{ch:section distributions on torus}, begins with a motivating analogy where we briefly discuss the importance of incorporating the geometry of the surface when defining the uniform distribution on a circle and a sphere. Then, in Section \ref{section area unofrm}, we explore the area-uniform distribution on the curved torus, using its intrinsic geometry. Moving forward, in Section \ref{section dependent model} we define a five-parameter dependent toroidal (bivariate circular) distribution by transforming the area-uniform distribution on the curved torus. The roles of different parameters are described by plotting the contours of the proposed density in Sections \ref{section plot density}.
 Section \ref{section estimation}, provides the maximum likelihood estimation (MLE) for the parameters of the bivariate joint probability density. 
 
 Model-based circular-circular regression, for example, proposed by  \cite{jha2018circular} and  \cite{kato2008circular}, are well-established in the literature. However, a more natural and appealing approach involves deriving regression through conditional expectation. So, Section \ref{section regression} begins with an introduction to a regression through the conditional expectation. 
%   We derive the marginals for $\Phi$ and $\Theta$. We find that the marginal for 
% $\Phi$ and  the marginal for 
% $\Theta$  both follows a Cardioid distribution. Hence, using that, we derive the conditional distribution for $\Phi$  given $\Theta=\theta$ and $\Theta$  given $\Phi=\phi$. Although the conditional distribution for $\Theta$  given $\Phi=\phi$ is unknown to us, we see that the conditional distribution for $\Phi$  given $\Theta=\theta$ also follows the Cardioid distribution. So, to predict a random variable 
% $\Phi$ given $\Theta=\theta$, we use here the conditional expectation $E[\Phi \mid \Theta = \theta]$. 
Unlike linear random variables, the mean direction obtained from the trigonometric moment can suitably replace the conditional expectation (moment) in this context.  

In Section \ref{ch:section sampling}, we have presented an extensive simulation finding.  First, in Section \ref{section random generate for bivariate}, we proposed an exact simulation method for generating random samples from the Cardioid distribution. Using this method, we generated random samples from the joint density of the area-uniform distribution on the curved torus. Through a repeated utilization of the simulation method from the Cardioid distribution we demonstrate how to generate samples from the proposed bivariate dependent dependent model when a marginal and conditional distribution both follows the Cardioid distribution.
From the simulation, it is observed that the proposed method of exact sampling of Cardioid distribution gives approximately $50\%$ improvement over the method proposed by   \cite{diaconis2013sampling} which
utilizes the acceptance-rejection sampling from the same.
In the next Section \ref{section parameter estimation simulated}, we have used the \textit{Nelder-Mead} algorithm to estimate the parameters of the simulated data from the proposed bivariate density and used it to obtain the estimated regression model. Before concluding in Section \ref{conclusion}, 
in Section \ref{section data analysis}, we conduct an extensive study of real-life medical science datasets, related to astigmatism resulting from cataract surgery which is described below.\\
 
\textbf{Motivating example of astigmatism data:}
A cataract is a condition where the lens of the eye becomes cloudy, leading to a partial or complete loss of vision. This cloudiness occurs due to changes in the lens over time, affecting its clarity. Cataract surgery involves removing the cloudy lens and replacing it with an artificial one to restore sight. Modern cataract surgery focuses on quick recovery and reducing surgical impact. Sometimes, cataracts can become dense and hard, making them more challenging to remove, despite technological advancements. A well-known method is small incision cataract surgery (SICS), where a small cut is made in the eye's outer layer, allowing access to the lens for removal and replacement.

In manual small incision cataract surgery (SICS), the incision typically ranges from 5.5 mm to 7 mm. When the lens's core, known as the nucleus, moves into the front chamber of the eye, it is removed through a tunnel using a combination of mechanical and fluid forces, a method called the irrigating vectis technique. For more information on this, refer to  \cite{srinivasan2009nucleus}. \cite{keener1995nucleus} introduced a variation of SICS using a snare made from a blunt needle and steel wire to divide and remove the nucleus. 
An advanced method called Phacoemulsification (PE) was developed by \cite{kelman2018phaco}, which uses ultrasonic vibrations to break up the lens, allowing it to be removed through a small incision. A further refinement, Torsional PE, uses oscillatory motion to minimize the impact during surgery. However, PE is less common in developing countries due to its high cost, longer learning curve, and limited effectiveness on dense cataracts. As a result, SICS remains the most widely used technique for cataract removal.

\begin{figure}[b]
\centering
\subfloat[]{%
  \includegraphics[trim= 80 80 80 80, clip,width=0.4\textwidth, height=0.4\textwidth]{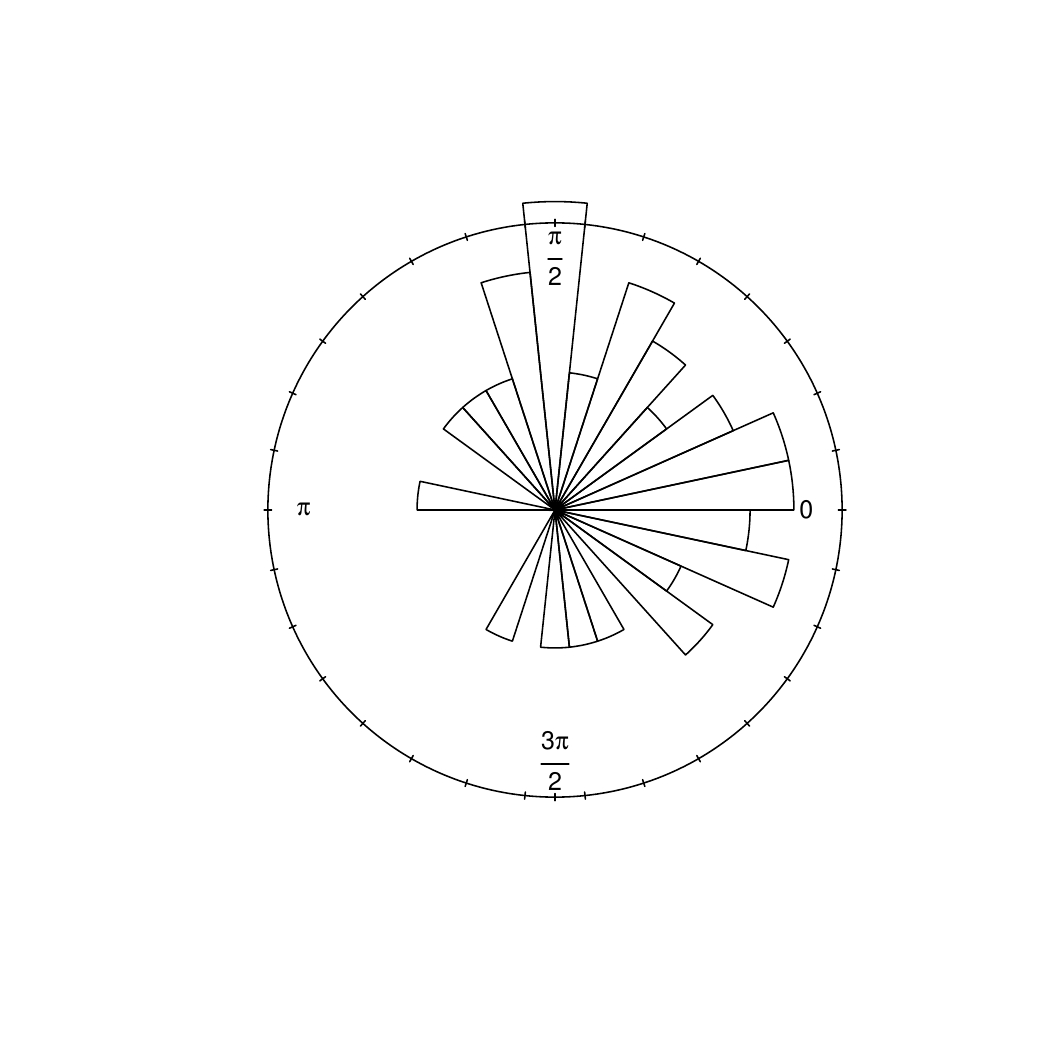}%
 
}
\subfloat[]{%
  \includegraphics[trim= 80 80 80 80, clip,width=0.4\textwidth, height=0.4\textwidth]{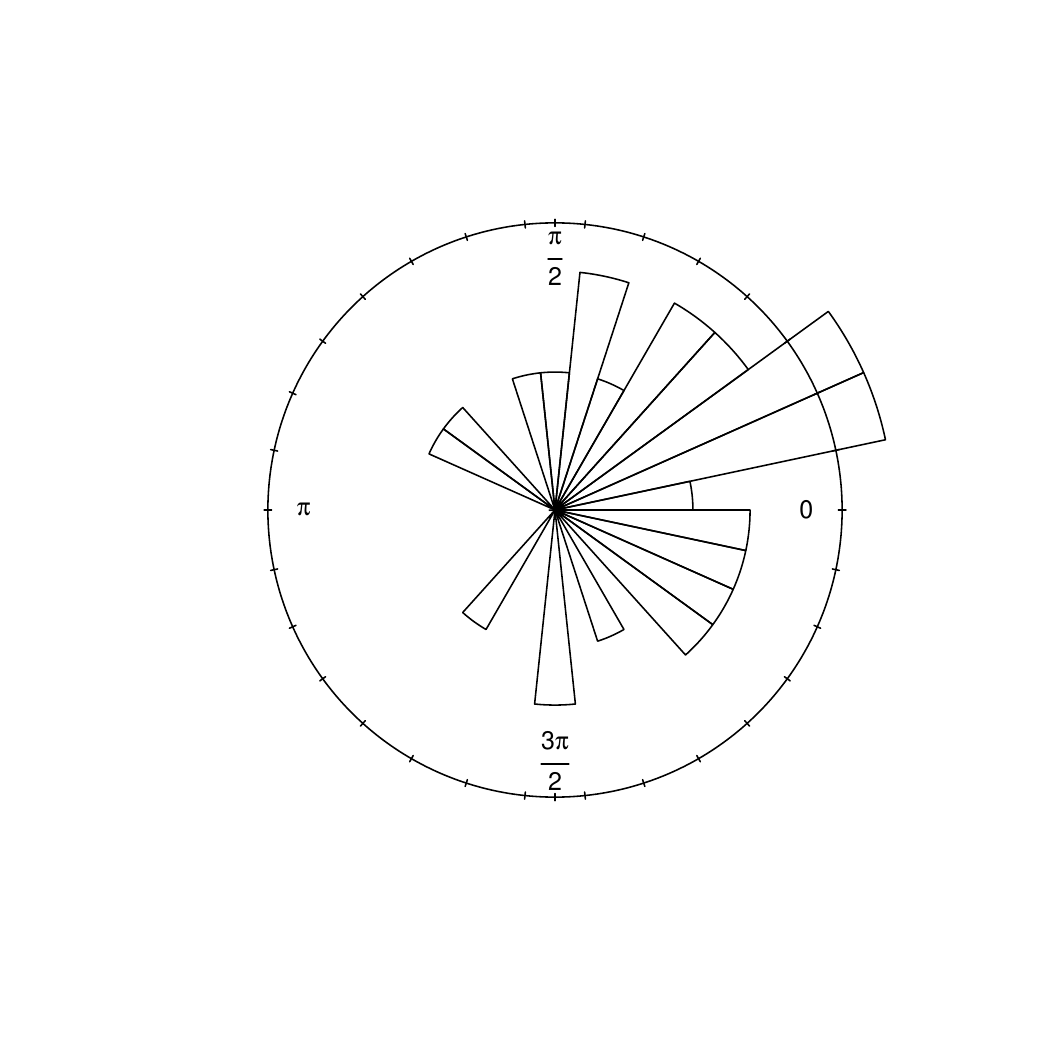}
  
}
\caption{(a) displays the rose-diagram of the axis of astigmatism (modulo $  360^{\circ}$) for the first month after surgery and (b) shows the same for the data after
third month of surgery. }
\label{subfig:figure rose plot}%
\end{figure}

During cataract surgery, the incision made in the cornea can sometimes alter its natural shape, leading to astigmatism. Astigmatism is a condition where vision becomes blurred because the eye cannot focus light onto the retina in a sharp, clear image. This occurs when the eye's curvature varies in different directions, causing different focal points. For instance, the image might be focused properly in one direction (like horizontally) but not in another (like vertically).
Astigmatism can be classified as regular or irregular, based on the orientation of the principal meridians of the eyes. In regular astigmatism, the two main meridians are perpendicular to each other, whereas, in irregular astigmatism, they are not. Regular astigmatism is further divided into three types:
\begin{enumerate}
    \item \textbf{With-the-rule astigmatism (WTR):} The vertical meridian is the steepest, similar to an American football lying on its side. Here, vertical lines appear clearer than horizontal lines.
    
    \item  \textbf{Against-the-rule astigmatism (ATR):} The horizontal meridian is the steepest, like an American football standing on its end. In this case, horizontal lines are seen more clearly.

    \item  \textbf{Oblique astigmatism:} The steepest curvature falls between angles of  $  120^{\circ}$ to  $  150^{\circ}$ or  $  30^{\circ}$ to $  60^{\circ}$. This type of astigmatism is more problematic because it distorts the vision of standard objects, which are usually oriented vertically or horizontally. As a result, oblique astigmatism can be more disorienting than WTR or ATR, where distortion occurs along more familiar axes.
\end{enumerate}
In general, it is preferable for the axis of astigmatism to be close to $0^{\circ}, 90^{\circ},$ or $  180^{\circ}$ for better visual outcomes. To align the preferred angle with the circular nature of the data, we multiply the angle by four ( mod $360^{\circ}$). As  
 \cite{mardia2000directional}, \cite{jammalamadaka2001topics} discuss, the typical method for converting axial data to circular data entails doubling the angle. However, in this particular scenario, doubling the angle once more is necessary to account for the preferred angles of  $0^{\circ}, 90^{\circ},$ or $  180^{\circ}$. For more detailed data processing one can see the article by  \cite{biswas2016comparison}.

A study conducted at the Disha Eye Hospital and Research Centre in Barrackpore, West Bengal, India, from 2008 to 2010 included 40 eyes from 40 patients (see  \cite{bakshi2010evaluation}). This was a comparative, prospective, randomized interventional trial. Patients were randomly assigned to one of two groups: 20 patients underwent small incision cataract surgery (SICS) with the snare technique, and 20 patients underwent SICS with the irrigating vectis technique. The follow-up period was three months. Figure-\ref{subfig:figure rose plot}(a) \& (b), displays the rose-diagram of the axis of astigmatism (mod $360^\circ$) for the first month after surgery and the third month after surgery, respectively. We represent the first-month follow-up axis of astigmatism data as the angle $\theta$, and the three months of the same as the angle $\phi$ for both surgical techniques. Together the pairs $(\phi_i,\theta_i)$  for $i=1, \cdots,40,$ represents a point on the  2-dimensional curved torus.

\section{Distributions on curved torus}\label{ch:section distributions on torus}
To comprehend the distribution of a random variable (vector) over a specified surface, it is crucial to understand the uniform distribution on that surface itself. A proper incorporation of the area element or Jacobian in the probability density function facilitates obtaining a uniform distribution on the surface.
For instance, when considering polar coordinates (circular data) for a circle with radius  $r$, then the area element (length element) is $dA_c=r~drd\theta$, and the Jacobian is $r.$ Consequently, the probability density function for a uniform distribution on the circle is given by
\begin{eqnarray}
    f(\theta)=\frac{r}{2\pi r }=\frac{1}{2\pi }\mbox{~for~} \theta \in [0,2\pi). 
\end{eqnarray}
Similarly, in the case of spherical coordinates (spherical data), where the area element is $ dA_s=r \sin\theta~ drd\theta d\phi$, and the Jacobian is $r \sin\theta$, the uniform distribution on the surface of the sphere with radius $r$ is 
\begin{eqnarray}
    f(\theta)=\frac{r \sin \theta}{4\pi r^2 }=\frac{1}{2\pi}\frac{\sin\theta}{2\pi r}\mbox{~for~} \theta \in[0,\pi),~ \phi \in [0,2\pi). 
\end{eqnarray}
The content in Ch. 9 of the book by \cite{mardia2000directional} extensively explored the unit spherical distribution with respect to the uniform distribution of spherical data. 
The geometry of circular and spherical data has been effectively incorporated into natural area-uniform distributions as well as related models. However, a similar geometric approach for the torus has rarely been explored, with the exception of the work by \cite{diaconis2013sampling}. Identifying the edges of a flat torus leads naturally to the constriction of a curved torus embedded in higher dimensions. Unlike a flat torus, a curved torus features all types of curvatures. This geometric structure allows for more natural modeling of the toroidal data, but at the same time, it demands a distinct way of analysis on curved torus. The rest of the paper is dedicated to that.

Now, we consider the  $2$-dimensional curved torus, a Riemannian manifold embedded in the $\mathbb{R}^{3}.$ Here, we will use the term ``curved torus'' for $2$-dimensional curved torus, that can be represented in parametric equations as
 \begin{equation}
 \begin{aligned}
      x(\phi,\theta) &= (R+r\cos{\theta})\cos{\phi}\\
    y(\phi,\theta) &= (R+r\cos{\theta})\sin{\phi} \\
    z(\phi,\theta) &= r\sin{\theta},
 \end{aligned}
 \label{torus equation}
 \end{equation}
   where  $R,r$ are radii of the horizontal and vertical circles, respectively.  The parameter space for the curved torus is $\mathcal{S}=\{ (\phi, \theta): 0\leq \phi, \theta<2\pi  \}$ which is commonly known as  ``flat torus''.

\subsection{Area-uniform toroidal distribution} \label{section area unofrm}
 Notably,  \cite{diaconis2013sampling} introduced an innovative method, for the first time in literature, to generate uniform random samples from the surface of a curved torus given by Equation-\ref{torus equation}.
They have defined uniformity by ensuring that random samples are drawn with frequencies proportional to the local area on the surface which is taken care of by the Jacobian of the curved torus.\\

Applying the principles from differential geometry (see Appendix), one can define the uniform distribution on the curved torus with respect to the area measure, and as a consequence, it will have the joint probability  density function \begin{equation}
   h^{*}(\phi,\theta)=\frac{1}{2\pi}\dfrac{(1+ \nu \cos{\theta})}{2\pi}, 
   \label{area-uniform dist}
\end{equation}
where $0\leq \phi,\theta <2\pi$ and $0 < \nu=\frac{r}{R} \leq 1.$ It is immediate from the above equation
that the horizontal angle (associated with $R$) with the marginal density  $ h^{*}_{1}(\phi)=\frac{1}{2\pi} \mbox{~where~~}  0\leq \phi<2\pi$ and the vertical angle (associated with $r$) with the marginal density \begin{equation}
   h^{*}_{2}(\theta)=\dfrac{(1+\nu \cos{\theta})}{2\pi}, \mbox{~where~~} 0\leq \theta<2\pi.  \label{Cardioid_pdf}
 \end{equation}
are independently distributed.
It is interesting to note that the above Equation-\ref{Cardioid_pdf} represents a Cardioid distribution.

\begin{figure}[t]
\centering
\subfloat[]{%
  \includegraphics[width=2 in]{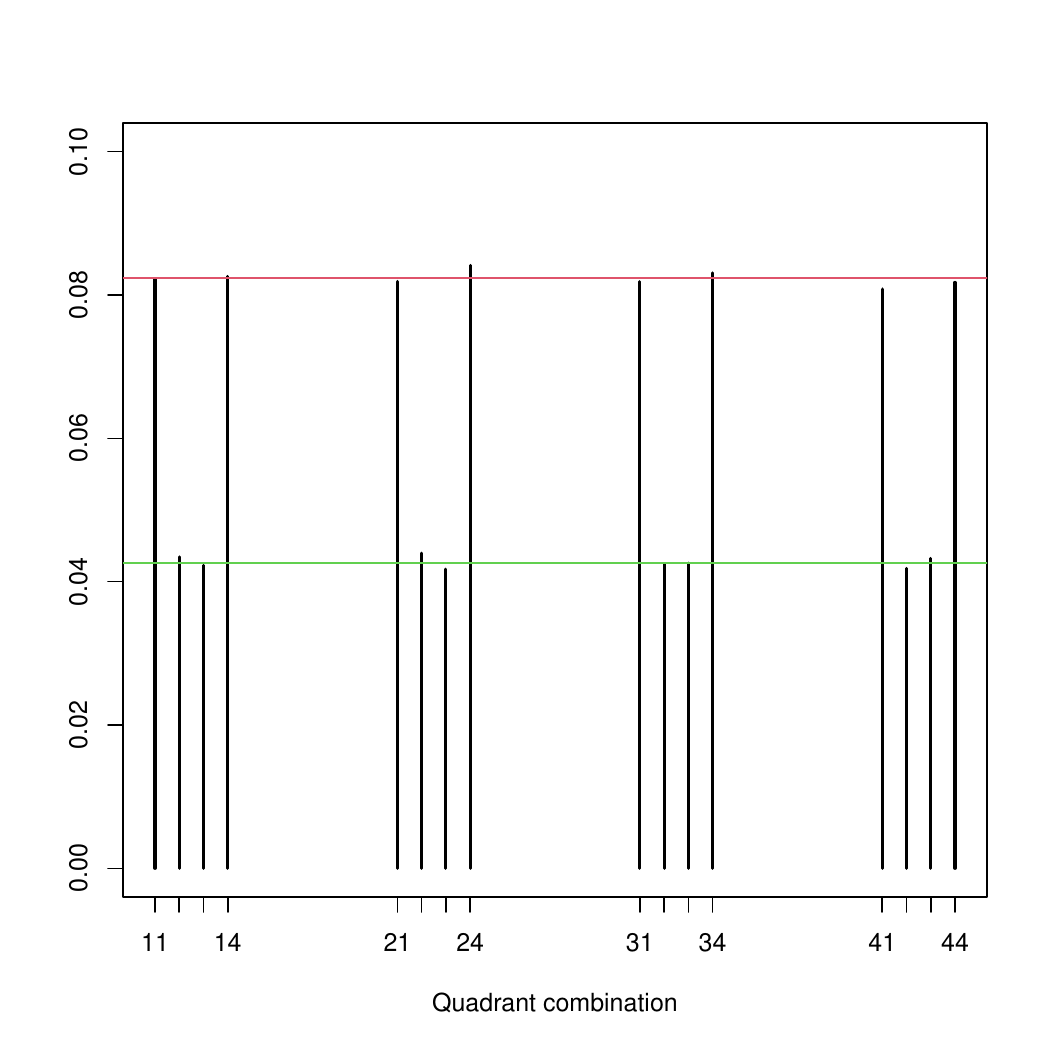}%
 
}
\subfloat[]{%
  \includegraphics[width=2 in]{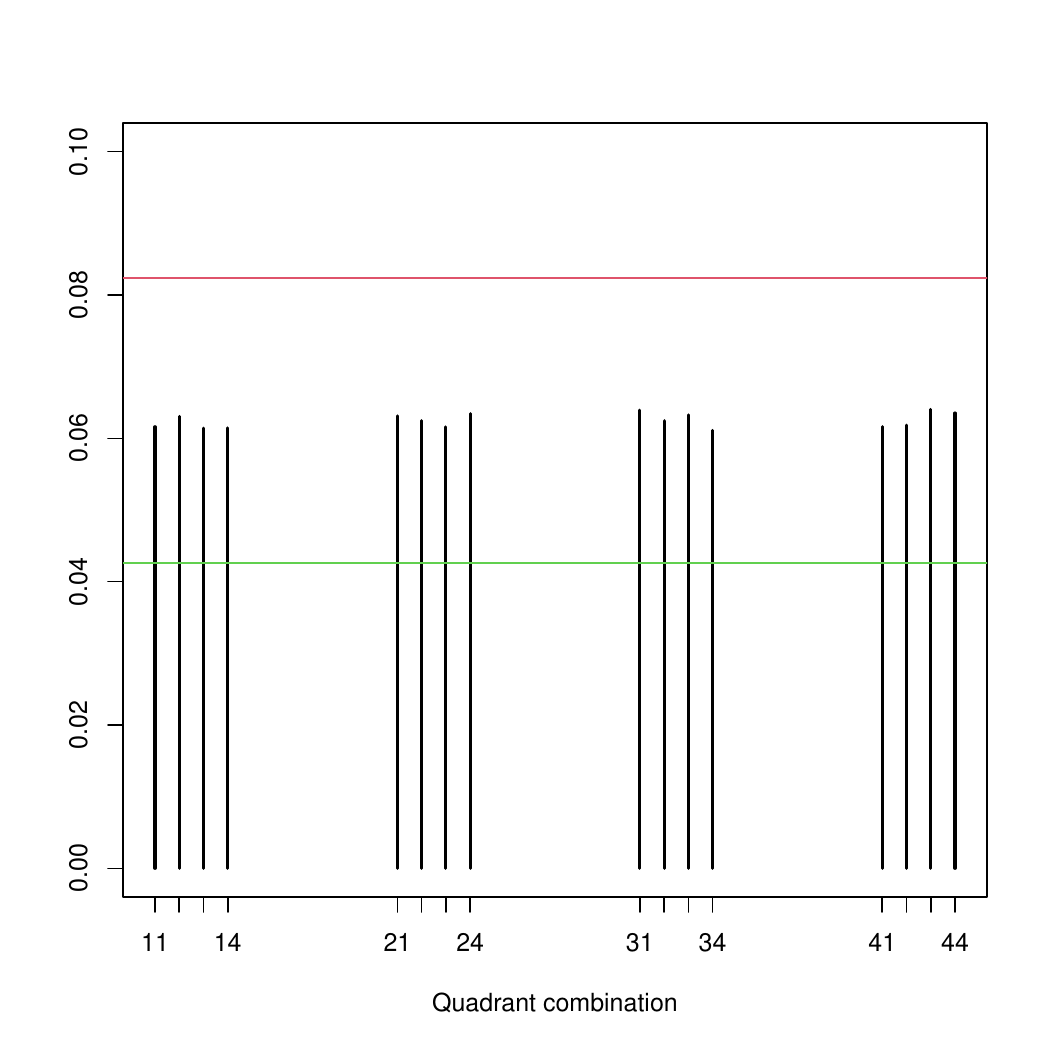}\hspace{2pt}
  
}\\
\subfloat[]{%
  \includegraphics[width=2 in]{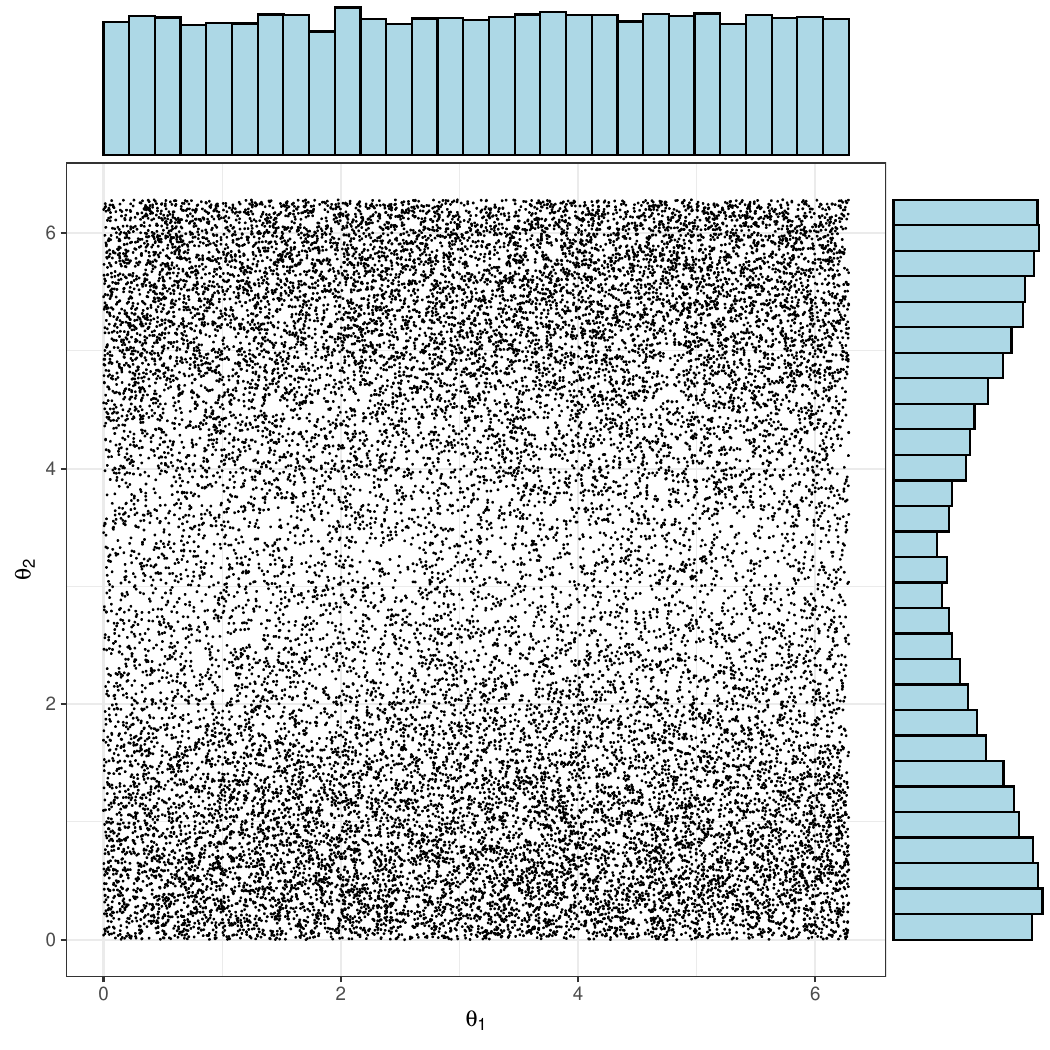}%
  
}
\subfloat[]{%
  \includegraphics[width=2 in]{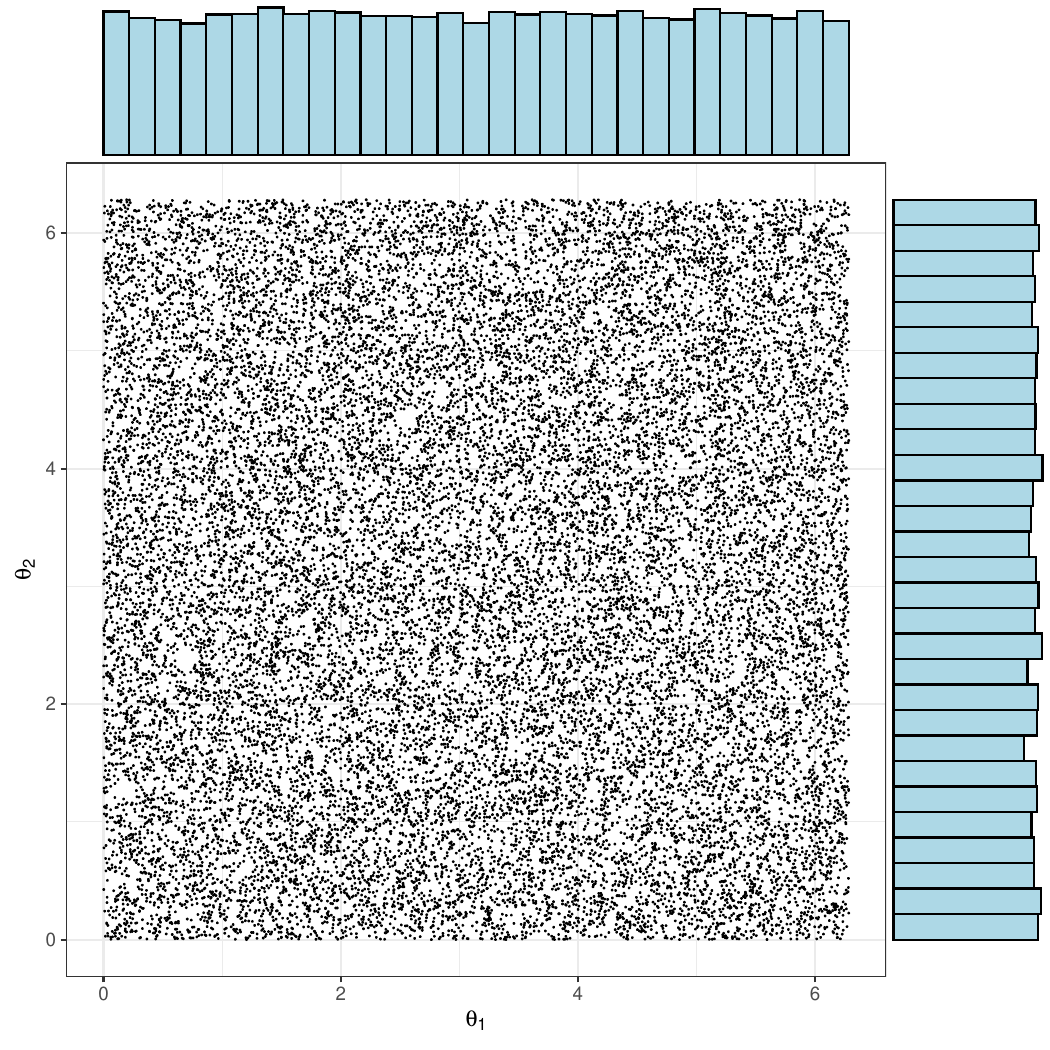}%
   
}
\caption{ (a) Bar diagram of the relative frequency table of uniformly distributed data on the surface of a curved torus drawn using Algorithm-\ref{algo_eau}. (b) Bar diagram of the relative frequency table of uniformly distributed data on the flat torus when projected on the curved torus. The red and green lines are the area proportionate to the positive and negative curvatures of the torus, respectively. (c) Scatterplot and the histogram of the marginal distributions of the uniformly distributed data on the surface of a curved torus drawn using Algorithm-\ref{algo_eau}. (d) Scatterplot and the histogram of the marginal distributions of the uniformly distributed data on the flat torus.}
\label{subfig:area_sampling}%
\end{figure}

 A pair of bar diagrams of relative frequency can make the difference between uniform samples from the flat and area-uniform samples from a curve torus.   We consider the quadrant combinations of horizontal and vertical circles as $Q_H\times Q_V$, with $Q_H=Q_V=\{1,2,3,4\}$, where each quadrant is of a partition with length $\pi/2.$ The bar diagram of Figure-\ref{subfig:area_sampling}(a) represents the relative frequency of uniformly generated points on the torus using the proposed sampling method. Whereas the bar diagram of Figure-\ref {subfig:area_sampling}(b) represents the relative frequency of generated points from the flat torus with the uniform distributions of angular parameters when projected on the curved torus. In both diagrams, the red and green lines represent the proportion of area to the quadrant combinations in positive and negative curvatures, respectively, to the total surface area of the curved torus. Although in  Figure-\ref{subfig:area_sampling}(a), the relative frequencies of quadrant combinations match with the respective proportions of the areas, in Figure-\ref{subfig:area_sampling}(b), it fails to do so. One can also be represented as the marginal plots of the data. Figure-\ref{subfig:area_sampling}(c) exhibits the marginal distribution of pre-image of the data uniformly drawn from the surface of a curved torus with respect to the area measure, whereas Figure-\ref{subfig:area_sampling}(d)  shows the similar plot of the data uniformly drawn from its parameter space, which is a flat torus.

\subsection{Dependent toroidal model} \label{section dependent model}
The aim of this subsection is to propose a dependent model for toroidal data arising from the area-uniform distribution on the curved torus. The joint probability density function of the proposed model  is given by
\begin{eqnarray}
h_{3}(\phi,\theta)=\frac{1}{4\pi^2} \big[1+ \nu \cos{(\theta-\mu_1)}\big]\big[1-\kappa \sin((\phi-\mu_2)+\lambda(\theta-\mu_1))\big],
\label{dependent pdf}
\end{eqnarray}

where $0\leq \phi,\theta <2\pi$, $0 < \nu \leq 1$, $0\leq \mu_1,\mu_2 <2\pi$, $-1 \leq \kappa \leq 1$, $\lambda \in \mathbb{R}$.
It is interesting to note that both the marginal densities are of Cardioid distribution. Moreover, one of the conditional distributions, $\Phi|\Theta=\theta$  is also a Cardioid distribution.

% \begin{theorem}
%   The the normalizing constant $C$ of the joint probability density function in Equation-\ref{dependent pdf}  is given by $~~~~C=4\pi^2.$
% \end{theorem}

% \begin{proof}
%     Consider 
%      \begin{eqnarray}
%         C &=& \int_{0}^{2\pi} \int_{0}^{2\pi} \big[1+ \nu \cos{(\theta-\mu_1)}\big]\big[1-\kappa \sin((\phi-\mu_2)+\lambda(\theta-\mu_1))\big]  \,dx \nonumber\\
%         &=& \int_{0}^{2\pi} \int_{0}^{2\pi}   \big[1+ \nu \cos{(\theta-\mu_1)}\big] \,d\phi ~d\theta \nonumber\\
%         &+& \kappa\int_{0}^{2\pi}    \big[1+ \nu \cos{(\theta-\mu_1)}\big] \bigg[ \cos((\phi-\mu_2)+\lambda(\theta-\mu_1))\bigg]_{0}^{2\pi} ~d\theta . \nonumber\\
%         &=& 2\pi *2\pi+0
%         \nonumber\\
%         C&=& 4\pi^2.
%         \nonumber
%     \end{eqnarray}
% \end{proof}
    
Depending upon the parameters, the distribution with the joint probability density function in Equation-\ref{dependent pdf} has the following special cases.

\begin{enumerate}
    \item If $\kappa=0$ or $\lambda=\frac{(n\pi-\phi-\mu_2)}{(\theta-\mu_1)}$, the joint probability density function in Equation-\ref{dependent pdf} becomes independent, with $\phi$ following a circular uniform distribution on $[0,2\pi]$ and $\theta$ following a Cardioid distribution with the mean direction at $\mu_1$. Now, if $\mu_1=0$, the joint probability density function in Equation-\ref{dependent pdf} becomes the area-uniform distribution on the curved torus, having the joint probability density function given in Equation-\ref{area-uniform dist}.

   \item If $\nu \rightarrow 0$, then the joint probability density function in Equation-\ref{dependent pdf} becomes
$$h_{4}(\phi,\theta)=\frac{1}{4\pi^2}\big[1-\kappa \sin((\phi-\mu_2)+\lambda(\theta-\mu_1))\big],$$  
where $0\leq \phi,\theta <2\pi$, $0\leq \mu_1,\mu_2 <2\pi$, $-1 \leq \kappa \leq 1$, $\lambda \in \mathbb{R}$.

\end{enumerate}

\subsection{Plot of the joint density} \label{section plot density}
\begin{figure}[t]
\centering
\subfloat[]{%
  \includegraphics[width=1.5 in]{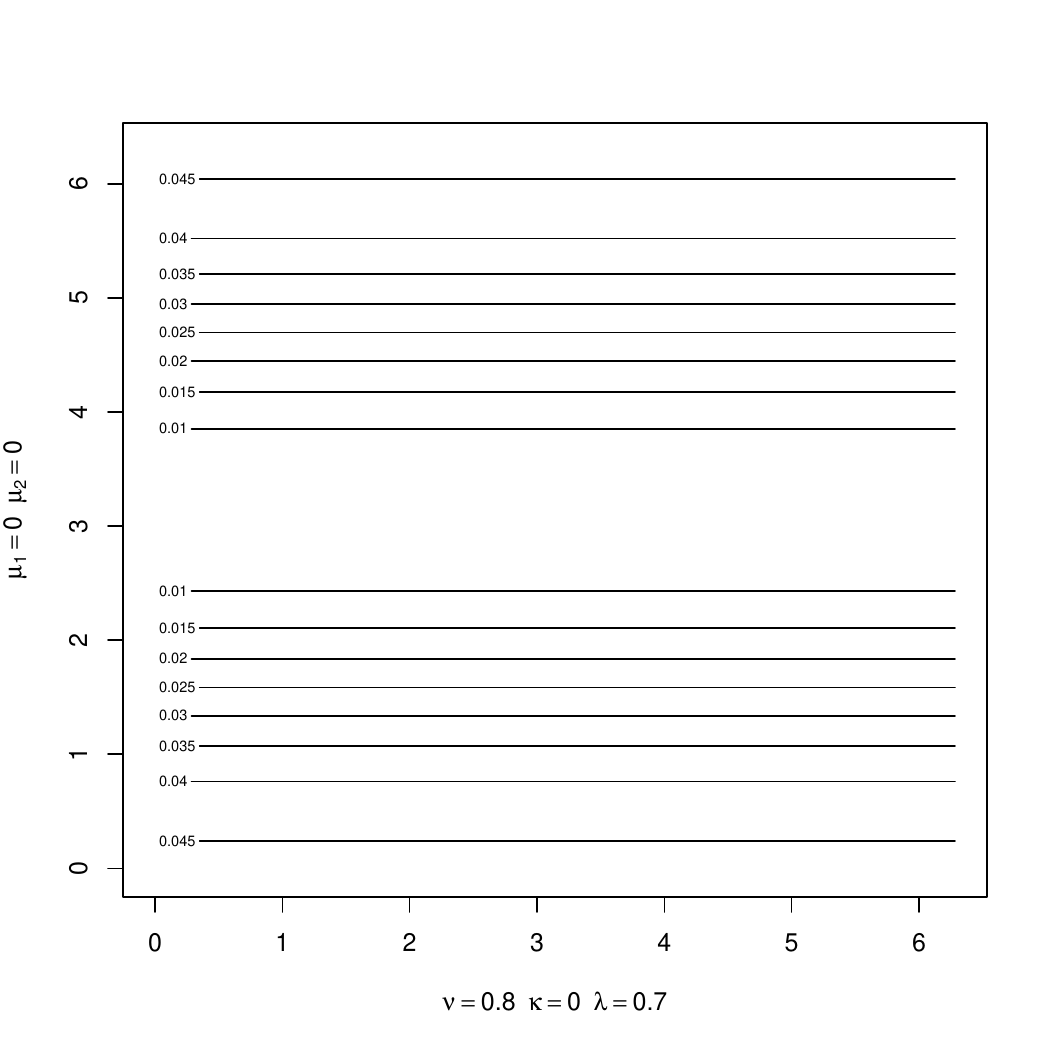}%
}
\subfloat[]{%
  \includegraphics[width=1.5 in]{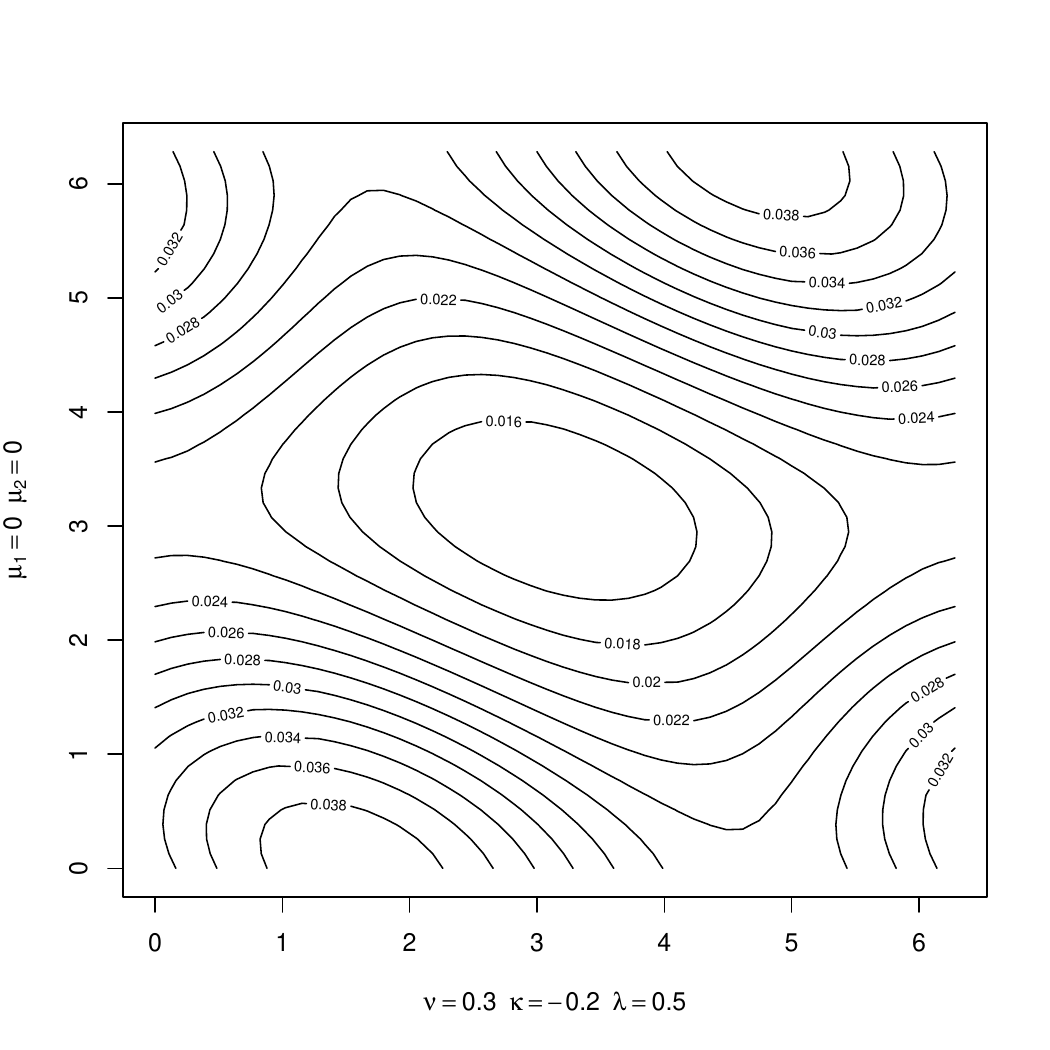}
}
\subfloat[]{%
  \includegraphics[width=1.5 in]{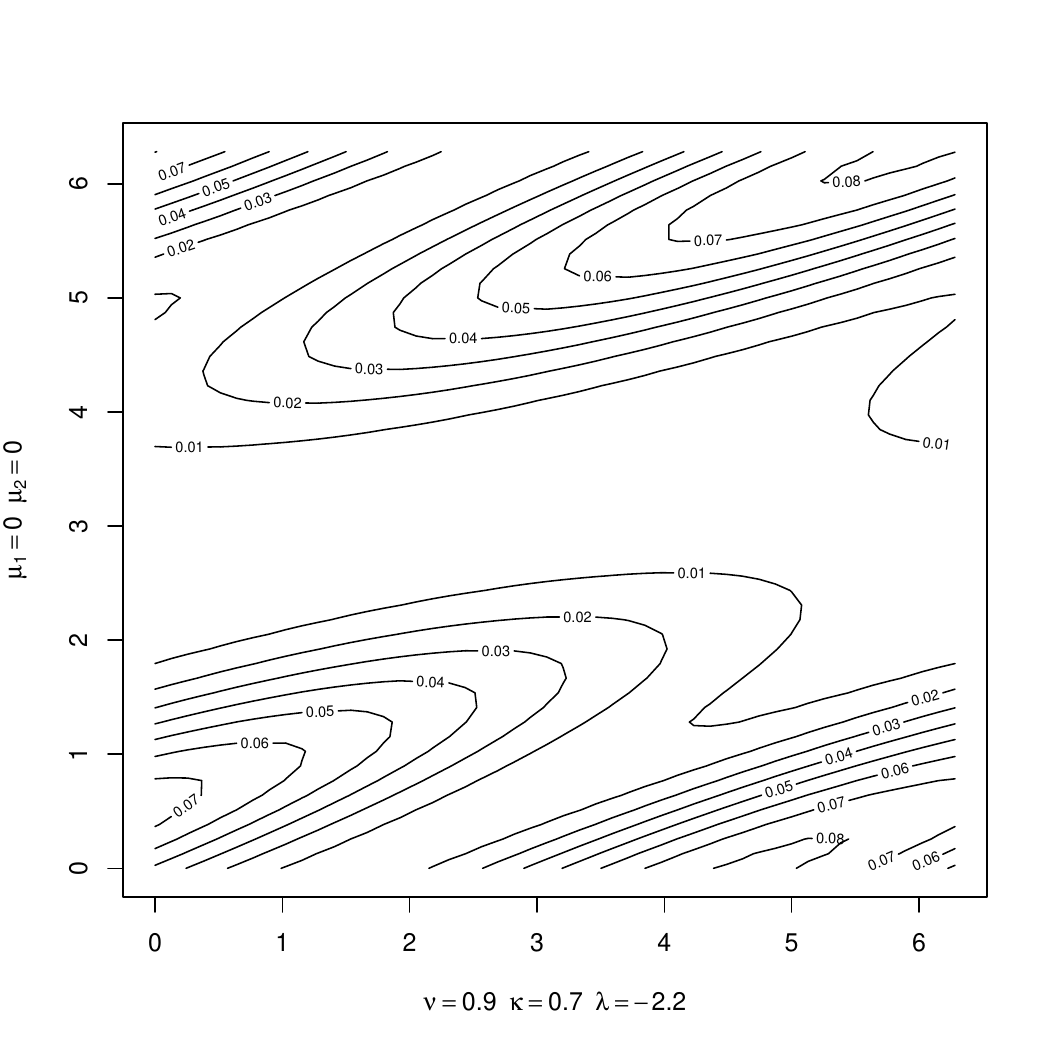} 
}
\subfloat[]{%
  \includegraphics[width=1.5 in]{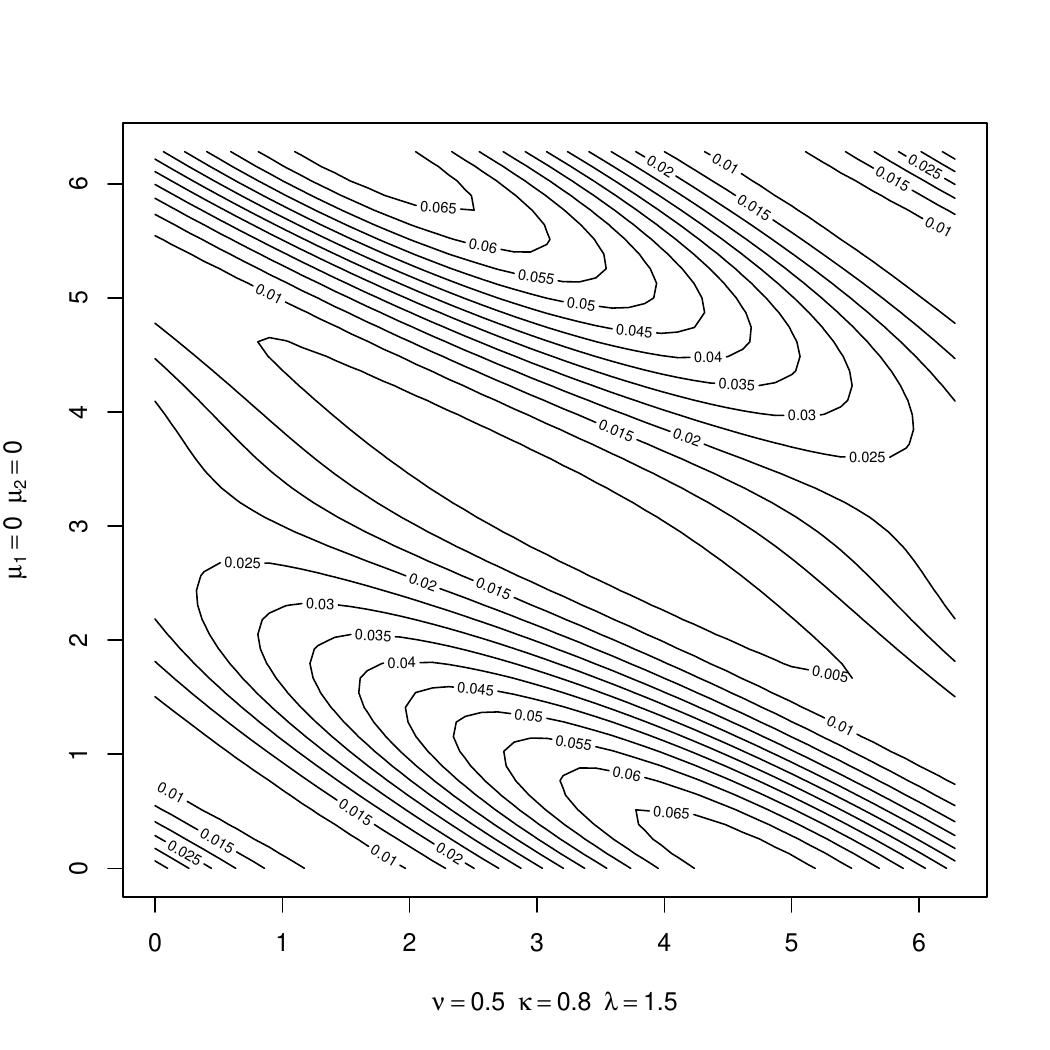}
}
\\
\subfloat[]{%
  \includegraphics[width=1.5 in]{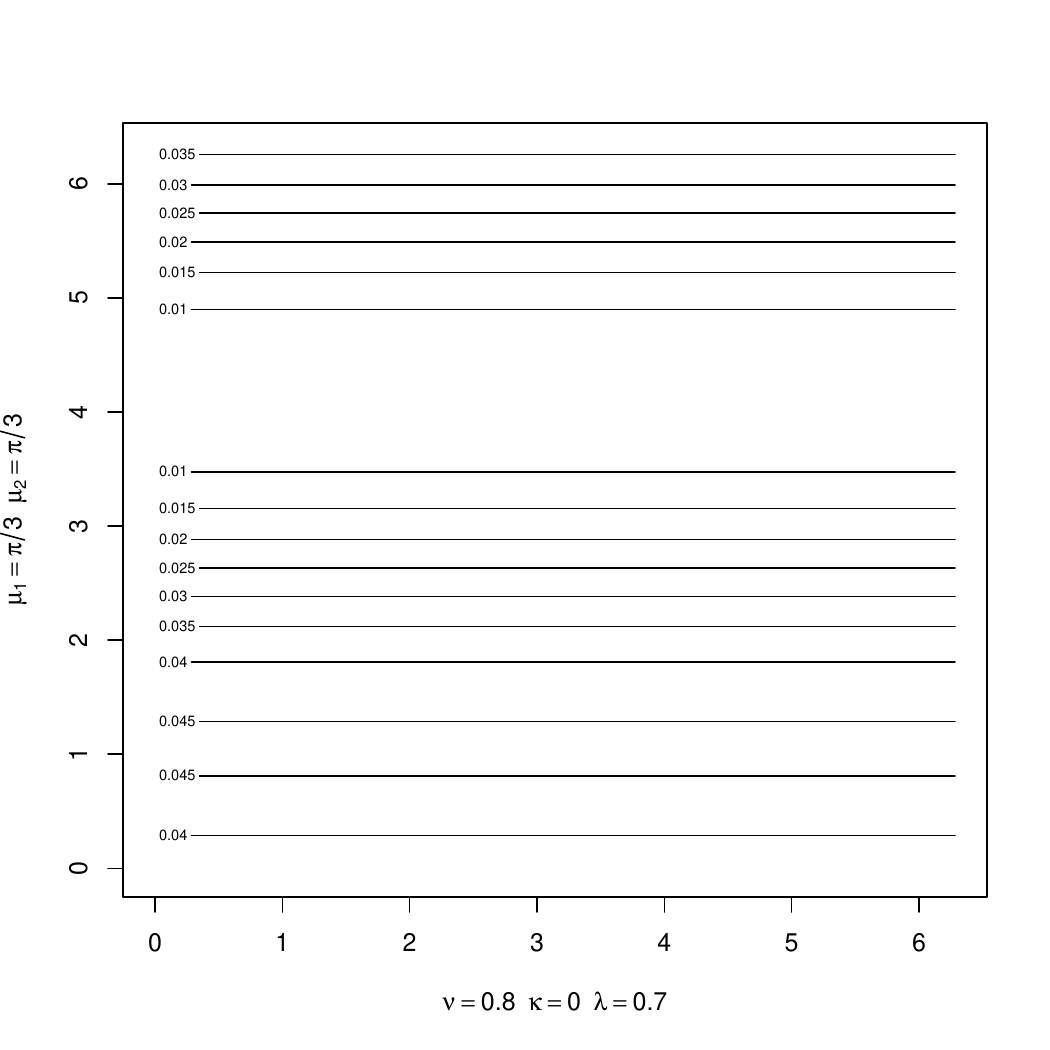}%
  
}
\subfloat[]{%
  \includegraphics[width=1.5 in]{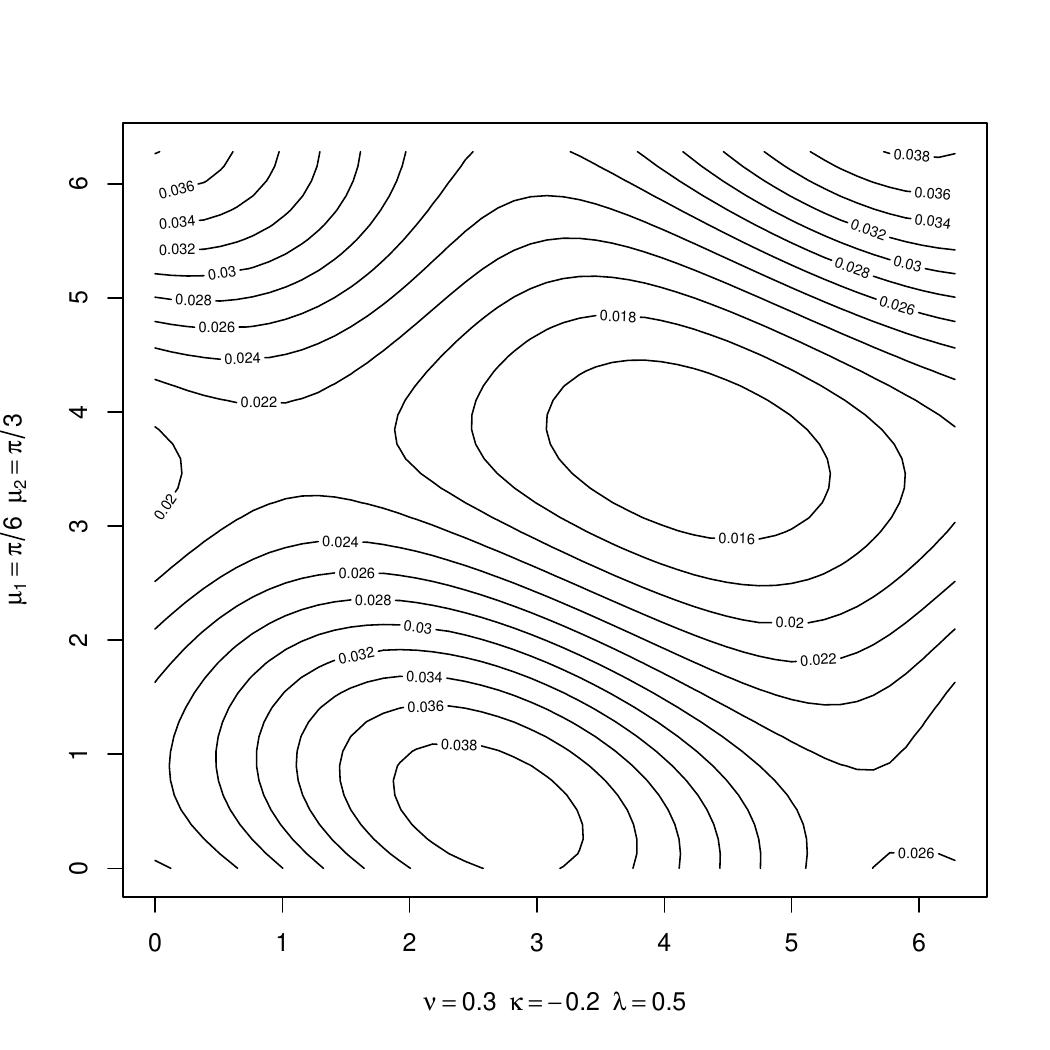}%
   
}
\subfloat[]{%
  \includegraphics[width=1.5 in]{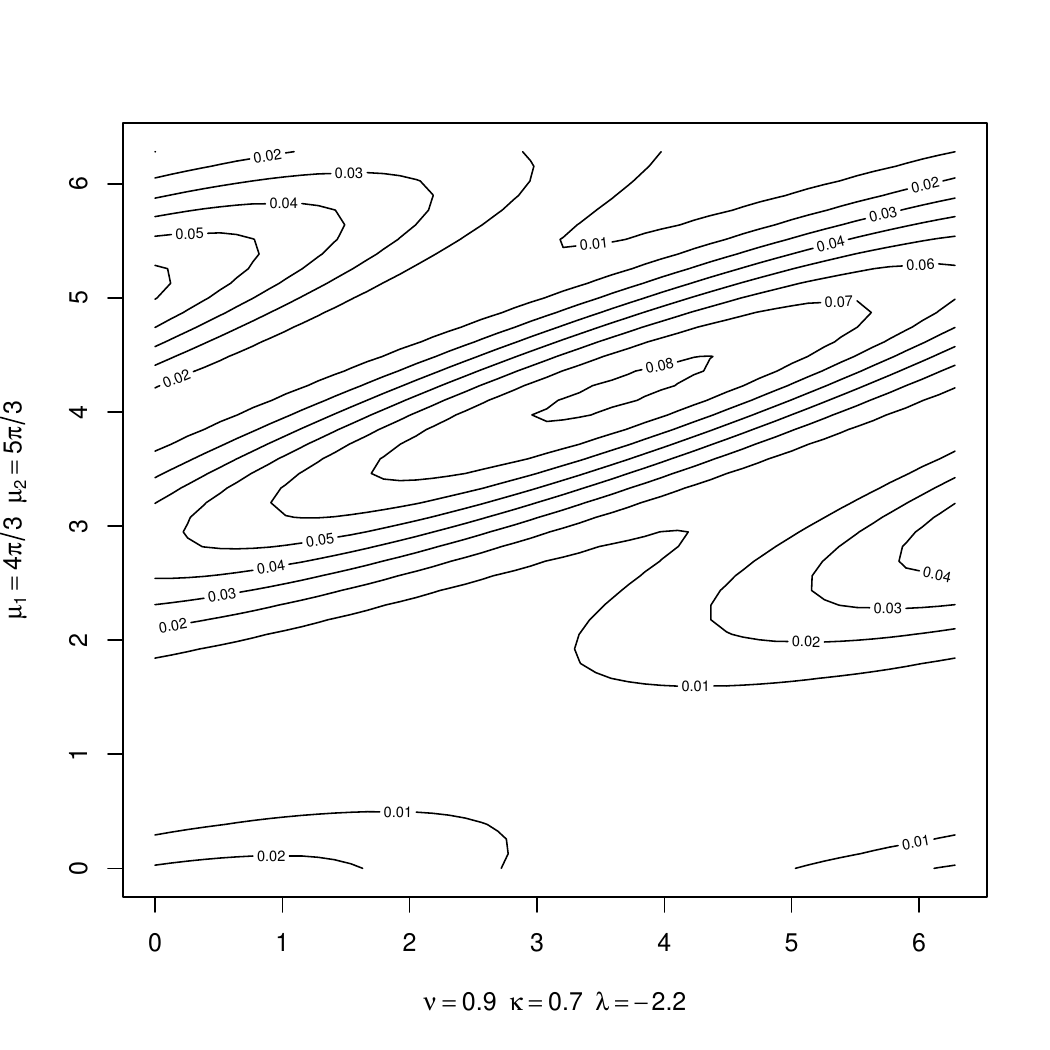}%
   
}
\subfloat[]{%
  \includegraphics[width=1.5 in]{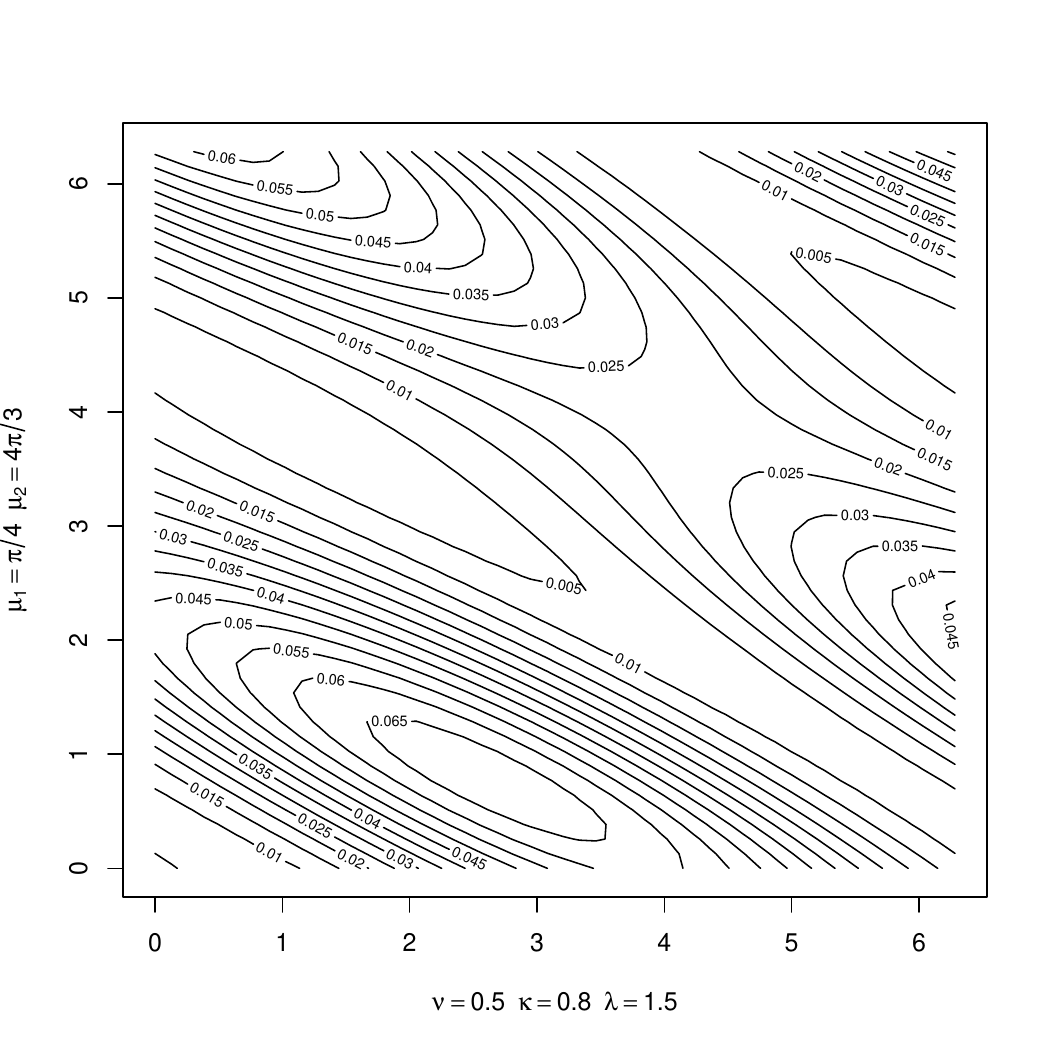}%
   
}
\caption{Contour plot of the bivariate proposed density for different parameters indicated in the labels. The values of the horizontal and vertical axes are the circular variables $\phi$ and $\theta$, respectively. In the top row, Figures-(a), (b), (c), and (d) show the effect of the concentration and dependence parameters for $\mu_1=\mu_2=0$, and in the bottom row, Figures-(e), (f), (g), and (h) display the same for non-zero $\mu_1$ and $\mu_2$.}
\label{subfig:bivariate_contur_plot_simulated}%
\end{figure}

The contour plot of the suggested bivariate density for various parameters is shown in Figure-\ref{subfig:bivariate_contur_plot_simulated}. For the top row, we have set the values of $\mu_1$ and $\mu_2$ to be 0. The other parameter values are as follows: for (a) $\nu=0.8,\kappa=0.0,\lambda=0.7$, for (b) $\nu=0.3,\kappa=-0.2,\lambda=0.5$, for (c) $\nu=0.5,\kappa=0.8,\lambda=1.5$, and for (d) $\nu=0.9,\kappa=0.7,\lambda=-2.2$. For the bottom row, (e) all other parameters remain unchanged from (a) and $\mu_1=\pi/3,\mu_2=\pi/3$, (f) all other parameters remain unchanged from (b) and $\mu_1=\pi/6,\mu_2=\pi/3$, (g) all other parameters remain unchanged from (c) and $\mu_1=\pi/4,\mu_2=4\pi/3$, (h) all other parameters remain unchanged from (d) and $\mu_1=4\pi/3,\mu_2=5\pi/3$.
From this illustration, it is evident that $\mu_1$ and $\mu_2$ influence the mean location of the data and have no impact on the concentration or the dependence structure of the data. The parameter $\kappa$ has a role as a concentration parameter, except when $\kappa=0$, while the parameter $\lambda$ determines the dependency structure.

\section{Model parameter estimation} \label{section estimation}
For a random sample $(\phi_1,\theta_1), \cdots, (\phi_n,\theta_n)$ drawn from the probability density function given in Equation- \ref{dependent pdf}. We intend to obtain the maximum likelihood estimators of the parameters.

The joint log-likelihood function is given by 
\begin{eqnarray}
    \mathcal{L}(\nu, \kappa, \lambda, \mu_1, \mu_2) &=& \sum_{i=1}^{n} \log \left( h_{3}(\phi_i, \theta_i) \right) \nonumber\\
    &=& \sum_{i=1}^{n} \log \left( \frac{1}{4\pi^2} \big[1+ \nu \cos{(\theta_i-\mu_1)}\big]\big[1-\kappa \sin((\phi_i-\mu_2)+\lambda(\theta_i-\mu_1))\big] \right) \nonumber\\
   &=& -2n\log(2\pi) + \sum_{i=1}^{n} \left( \log\left[1+ \nu \cos{(\theta_i-\mu_1)}\right] \right. \nonumber \\
   && \qquad\quad + \left. \log\left[1-\kappa \sin((\phi_i-\mu_2)+\lambda(\theta_i-\mu_1))\right] \right)
   \label{log_likelihood}
\end{eqnarray} 

Taking derivatives of Equation-\ref{log_likelihood} with respect to $\nu, \kappa, \lambda, \mu_1, \mu_2$ and equate with $0$ we get

\begin{eqnarray}
\frac{\partial \mathcal{L}}{\partial \nu} = \sum_{i=1}^{n} \left( \frac{\cos(\theta_i - \mu_1)}{1 + \nu \cos(\theta_i - \mu_1)} \right)=0
\end{eqnarray}

\begin{eqnarray}
\frac{\partial \mathcal{L}}{\partial \kappa} = \sum_{i=1}^{n} \left( \frac{-\sin((\phi_i - \mu_2) + \lambda (\theta_i - \mu_1))}{1 - \kappa \sin((\phi_i - \mu_2) + \lambda (\theta_i - \mu_1))} \right)=0
\end{eqnarray}

\begin{eqnarray}
\frac{\partial \mathcal{L}}{\partial \lambda} = \sum_{i=1}^{n} \left( \frac{-\kappa (\theta_i - \mu_1) \cos((\phi_i - \mu_2) + \lambda (\theta_i - \mu_1))}{1 - \kappa \sin((\phi_i - \mu_2) + \lambda (\theta_i - \mu_1))} \right)=0
\end{eqnarray}

\begin{eqnarray}
\frac{\partial \mathcal{L}}{\partial \mu_1} = \sum_{i=1}^{n} \left( \frac{-\nu \sin(\theta_i - \mu_1)}{1 + \nu \cos(\theta_i - \mu_1)} - \frac{\kappa \lambda \cos((\phi_i - \mu_2) + \lambda (\theta_i - \mu_1))}{1 - \kappa \sin((\phi_i - \mu_2) + \lambda (\theta_i - \mu_1))} \right)=0
\end{eqnarray}

\begin{eqnarray}
    \frac{\partial \mathcal{L}}{\partial \mu_2} = \sum_{i=1}^{n} \left( \frac{-\kappa \cos((\phi_i - \mu_2) + \lambda (\theta_i - \mu_1))}{1 - \kappa \sin((\phi_i - \mu_2) + \lambda (\theta_i - \mu_1))} \right)=0
\end{eqnarray}

The system of equations presented above lacks a closed-form solution due to the complexity of the functions involved. Consequently, numerical optimization techniques are recommended for parameter estimation.

\section{Regression} \label{section regression}
The predictability of the random variable $\Phi$ given $\Theta = \theta$ can be obtained through its conditional expectation i.e. $E[\Phi \mid \Theta = \theta]$. Although the model-based analysis of circular-circular regression is well studied in the literature, regression obtained through the conditional expectation from the joint density of the pair of bivariate circular random variables is more natural in this context.
To get $E[\Phi \mid \Theta = \theta]$, we first need to determine the 
corresponding marginal and conditional probability density.

Although for the linear random variables, regression is obtained through
the conditional expectation of one variable given the other, the same is not applicable for the circular random variables. Here the conditional expectation (moment) can be suitably replaced by the mean direction obtained from the trigonometric moment.

\begin{theorem}
The mean direction of the conditional probability density function of $ \Phi $ given $\Theta=\theta$ is
$$E(\Phi|\Theta=\theta)=\left[\frac{3\pi}{2}+\mu_2+\lambda(\theta-\mu_1)\right] \mod 2\pi $$ where $0\leq \theta <2\pi$,  $0\leq \mu_1,\mu_2 <2\pi$,  $\lambda \in \mathbb{R}.$
\label{coditional mean direction thm}
\end{theorem}

\begin{proof}
See in Appendix

\end{proof}

\begin{theorem}
    The conditional probability density function of $ \Theta $ given $\Phi=\phi$ is  
$$ h_6(\theta|\phi)=\frac{ \frac{1}{4\pi^2}  \big[1+ \nu \cos{(\theta-\mu_1)}\big]\big[1-\kappa \sin(\phi-\mu_2+\lambda (\theta-\mu_1))\big] }{  \frac{1}{2\pi}\left[1+A \cos \left(\phi-\mu_3\right) \right]}$$
where $0\leq \phi,\theta <2\pi$,  $0 < \nu \leq 1,$ $\mu_3=\left(\frac{3\pi}{2}-\lambda \pi \right)$ $ \mod 2\pi$, $0\leq \mu_1,\mu_2,\mu_3 <2\pi$, $-1 \leq \kappa \leq 1$, $-1 \leq A=\left[\frac{\kappa \{\lambda^2(1+\nu)-1\}}{\pi(\lambda^3-\lambda)} \sin{(\lambda\pi)} \right]\leq 1$, $\lambda \in \mathbb{R}.$
\label{conditional of theta thm}
\end{theorem}

% \begin{lemma}
%   The marginal probability density function of  $\Phi$   is given by
% $$h_5(\phi)=\frac{1}{2\pi} \left[1 - A \sin(\phi+\lambda\pi)\right], $$ where $A=\left[\frac{\kappa \{\lambda^2(1+\nu)-1\}}{\pi(\lambda^3-\lambda)} \sin{(\lambda\pi)} \right]$, $0\leq \phi <2\pi$, $0 < \nu \leq 1$, $-1 \leq \kappa \leq 1$, $\lambda \in \mathbb{R},$
% \label{marginal of phi}
% \end{lemma}

\begin{proof}
See in Appendix

\end{proof}

\begin{figure}[t]
\centering
\subfloat[]{%
  \includegraphics[width=2.2 in]{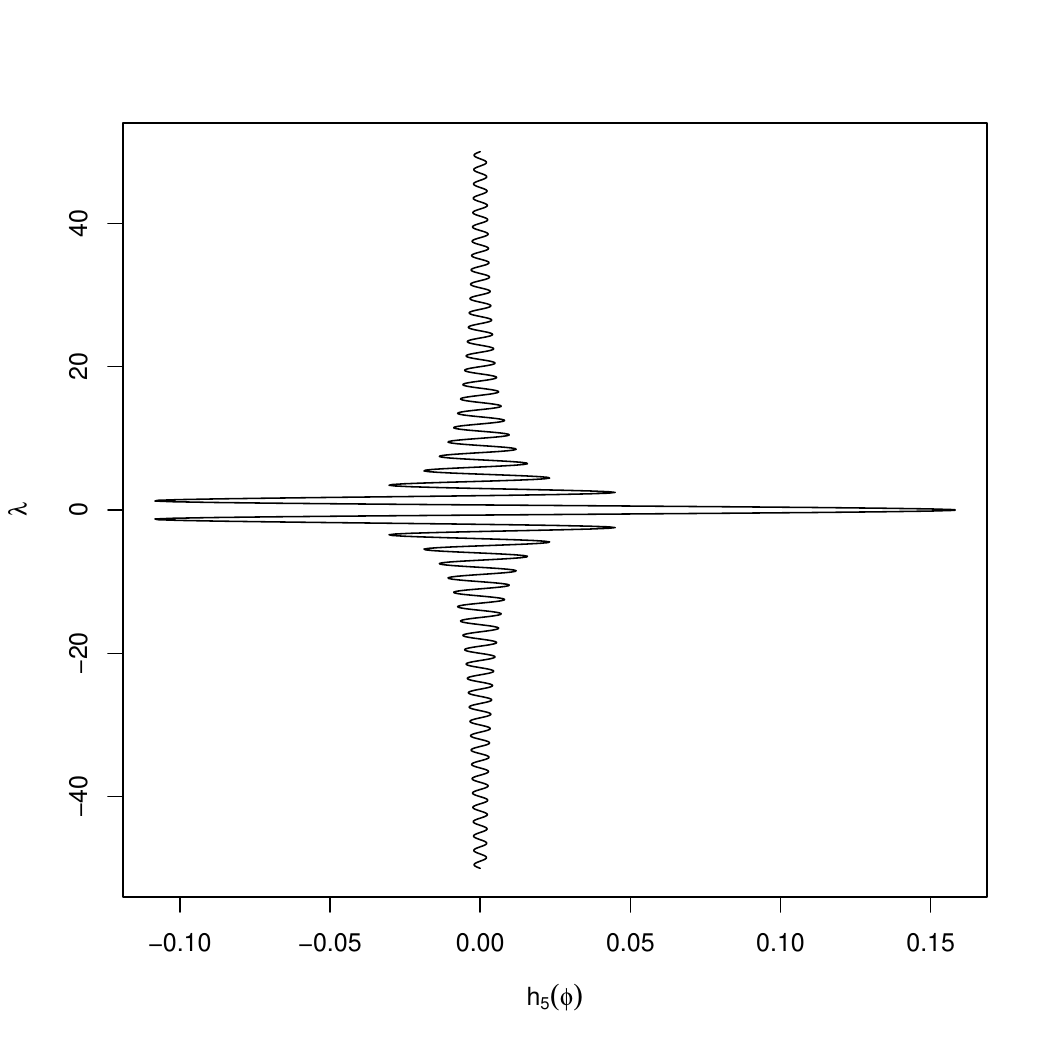}%
 
}
\subfloat[]{%
  \includegraphics[width=2.2 in]{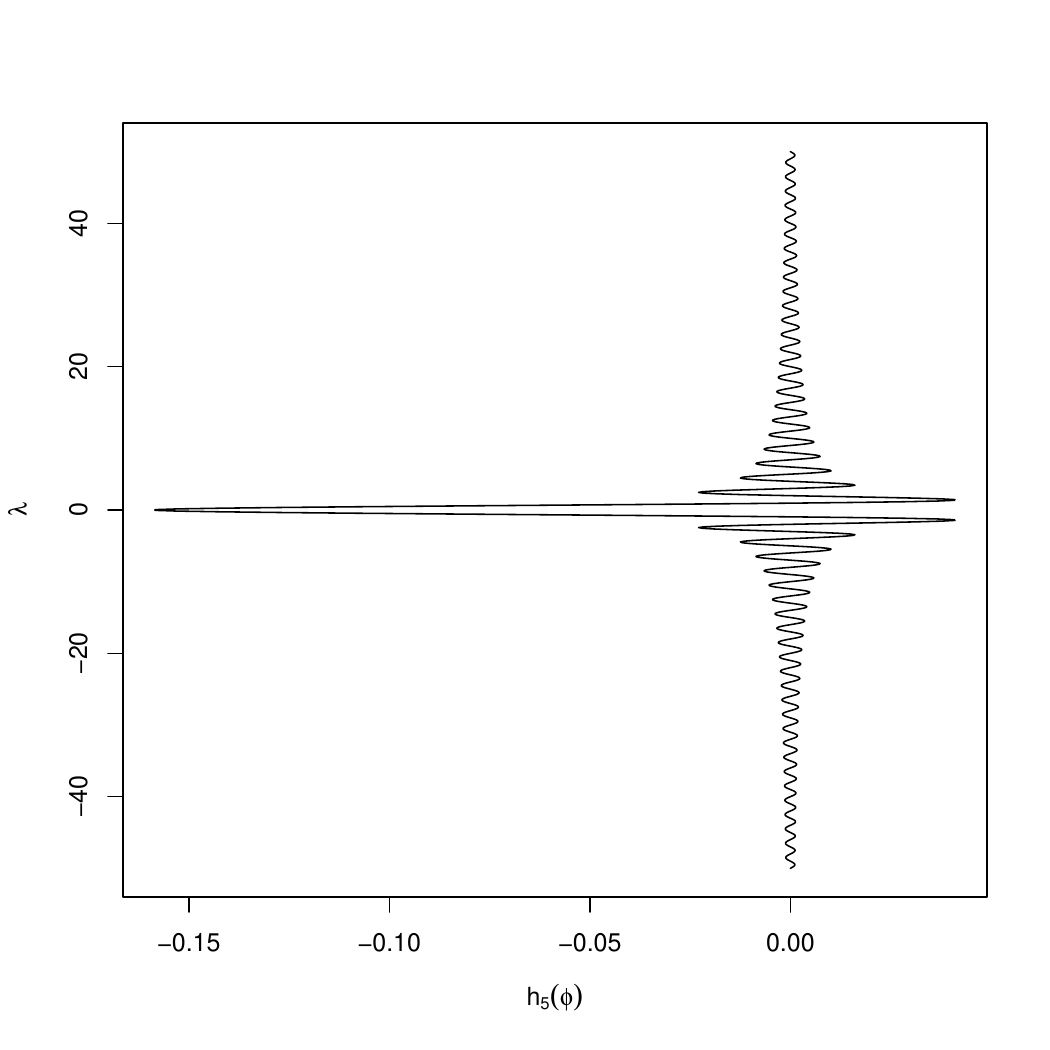}\hspace{2pt}
  
}
\subfloat[]{%
  \includegraphics[width=2.2 in]{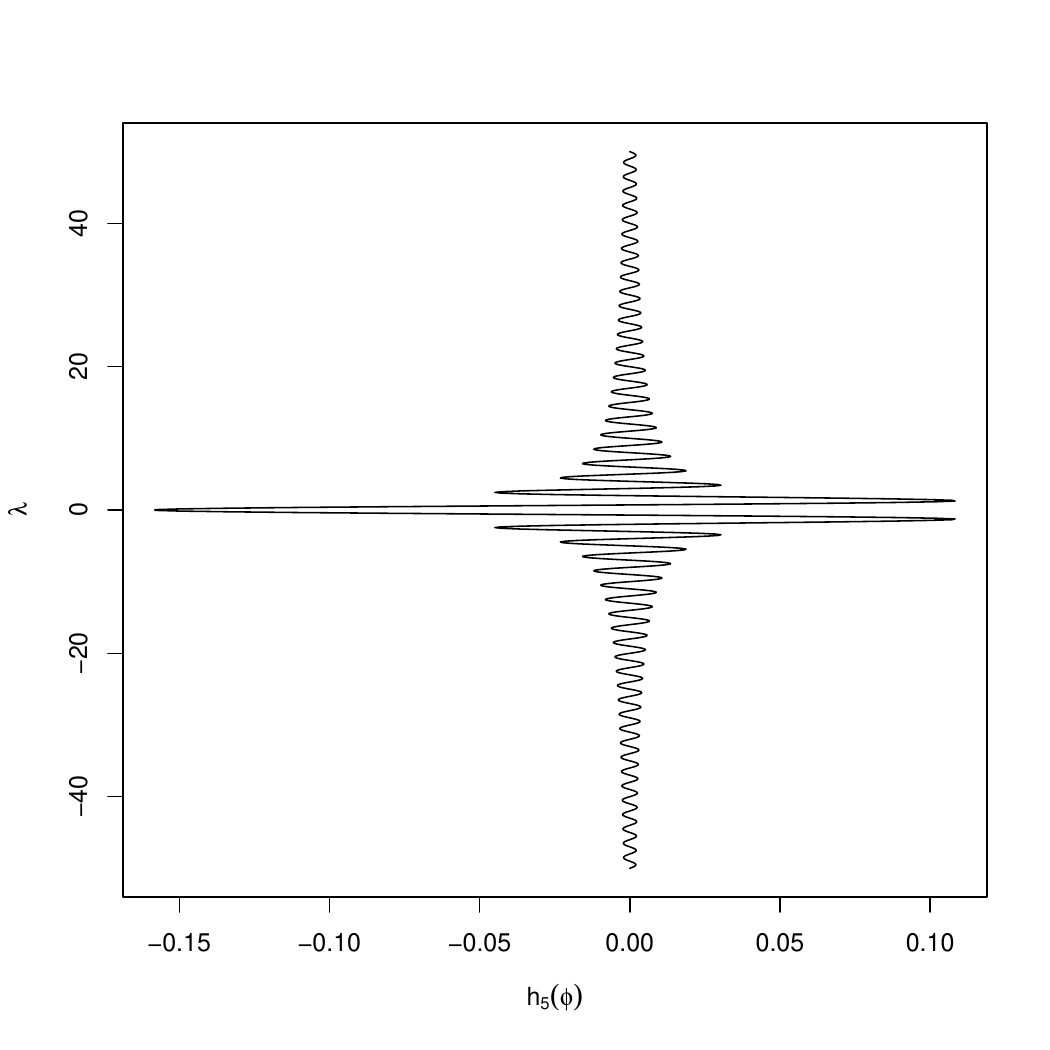}%
  
}\\
\subfloat[]{%
  \includegraphics[width=2.2 in]{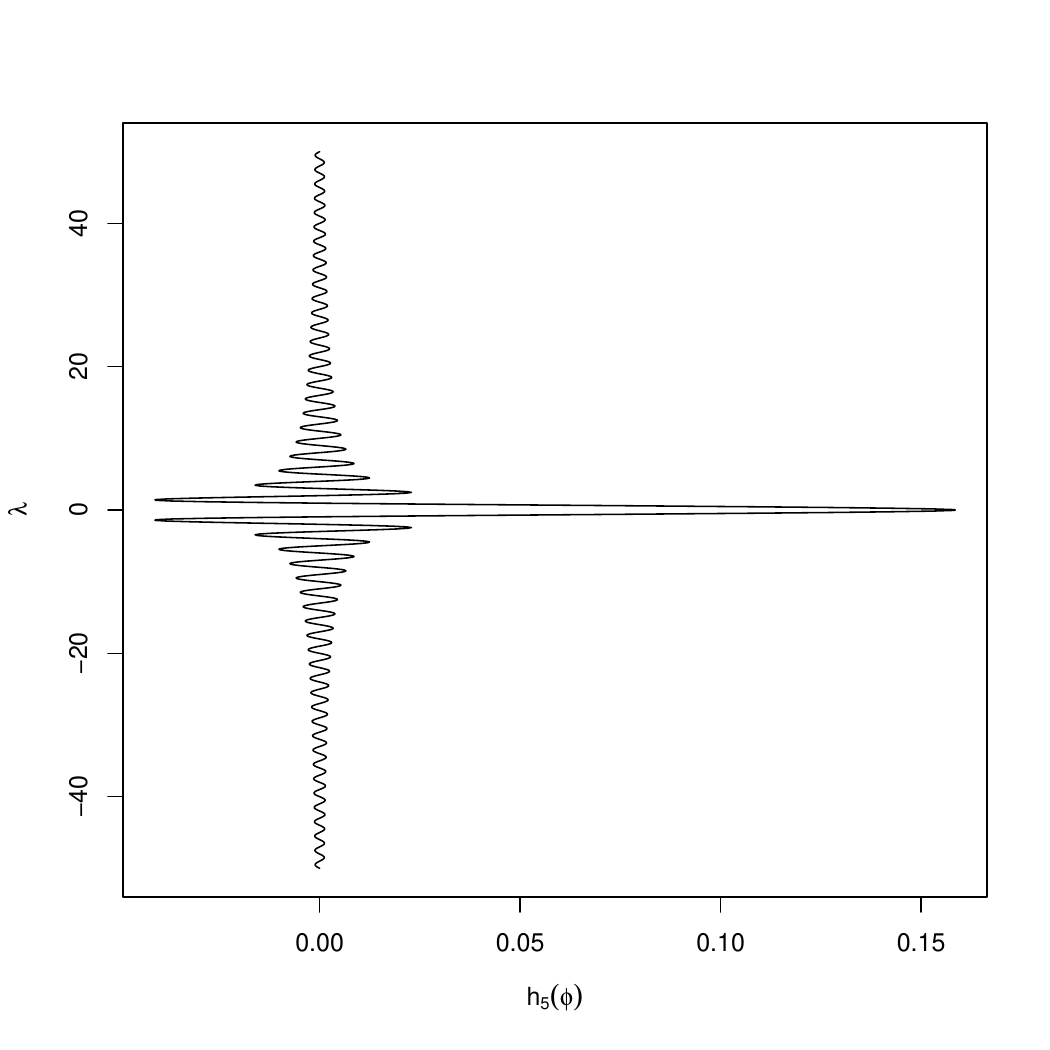}%
   
}
\subfloat[]{%
  \includegraphics[width=2.4 in]{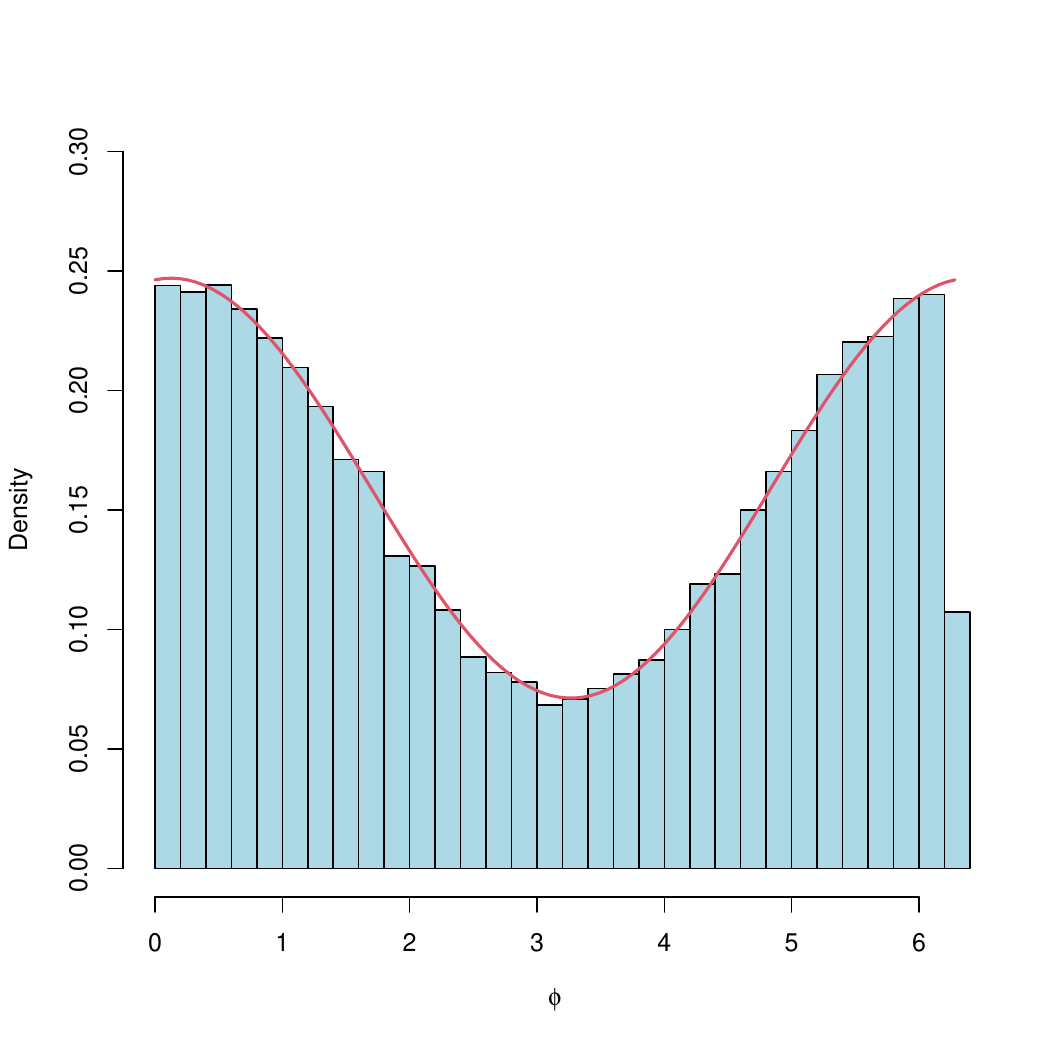}%
   
}
\subfloat[]{%
  \includegraphics[width=2.4 in]{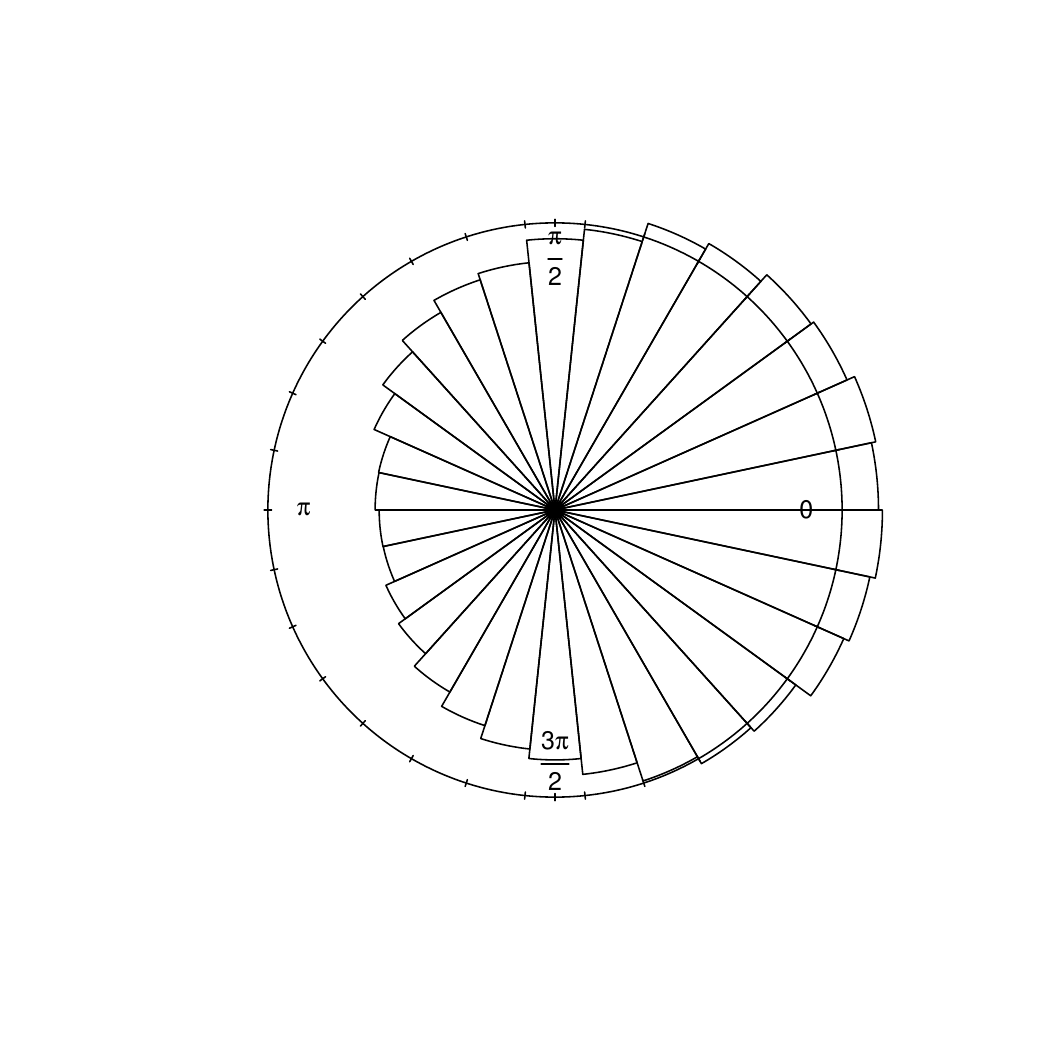}%
   
}
\caption{ The plot of $A=\left[\frac{\kappa \{\lambda^2(1+\nu)-1\}}{\pi(\lambda^3-\lambda)} \sin{(\lambda\pi)} \right]$ of the marginal distribution of $\Phi$ for $1000$ equispaced values of $\lambda$ taken from $-50$ to $50$ is shown. (a) Displays the plot for $\kappa=1$ and $\nu=1$. (b) Displays the plot for $\kappa=-1$ and $\nu=0.1$. (c) Displays the plot for $\kappa=-1$ and $\nu=1$. (d) Displays the plot for $\kappa=1$ and $\nu=0.1$. (e) shows the histogram of the sample of the marginal density, (f) is the corresponding rose plot.}
\label{subfig:marginal phi concentration plot}%
\end{figure}

\begin{rmk}

    One can observe that Figures-\ref{subfig:marginal phi concentration plot}(a), (b), (c), and (d) illustrate that $-1 \leq A \leq 1$ for different values of $\nu$ and $\kappa$ when $1000$ equispaced values of $\lambda$ taken from $-50$ to $50$. This behavior can also be demonstrated theoretically, although we omit this proof for brevity. Taken together, these observations allow us to conclude that the density in Equation-\ref{phi marginal cal2} corresponds to the density of a Cardioid distribution. Figure-\ref{subfig:marginal phi concentration plot}(e) and (f) depicts the histogram of this marginal distribution and rose plot, respectively, for $\nu=0.2, \kappa=-0.85,\lambda=0.46, \mu_1=0$, and $\mu_2=0$, when the sample is genereted using the Algorithm-\ref{algo_eau}.

\end{rmk}

\section{Simulation}\label{ch:section sampling}

This section will initially address the method of generating random variates from the Cardioid distribution. This is one of the marginal distributions with the probability density function $h^{*}_{2}(\theta)$ in Equation-\ref{Cardioid_pdf} of the area-uniform distribution of the torus.  The random sample from the bivariate joint probability density function in Equation-\ref{dependent pdf} is obtained by applying the sampling from Cardioid and the conditional distribution of $\Phi$ given $\Theta=\theta$. Finally, using the random samples, we discuss the parameter estimations via simulation.

\subsection{Exact sampling from Cardioid distribution} \label{section random generate for bivariate}
Here,  we propose an exact sampling method using a probabilistic transformation to generate samples from $h^{*}_{2}(\theta)$, described in the Theorem-\ref{EAU_thm}. The Algorithm-\ref{algo_eau} is the pseudo-code for the newly proposed exact sampling method, which is not an inversion of the cumulative distribution function. It can be noted that the Algorithm-\ref{algo_eau} is applicable for $\nu\in (-1,0)$ also.

We report the results of a thorough simulation study to compare the proposed exact sampling scheme with the existing area-uniform rejection (AUR) sampling algorithm provided by  \cite{diaconis2013sampling}. We consider different values of $\nu=\frac{r}{R}$ form $0.1$ to $1$ with equal gaps and compute the acceptance percentages for the sample size of $10000$ in both algorithms. The following Table-\ref{table:eau uni} shows that the proposed sampling outperforms AUR sampling to a large extent.

\begin{theorem}
    Suppose $U$ follows  uniform distribution on $[0, 1]$, $\Theta$ follows  uniform distribution on $[0, 2\pi]$, and $p(\Theta)=\dfrac{1}{2}(1+\nu\cos \Theta)$. Then the random variable $Y$ is defined by \\
 
 $Y = \left\{
        \begin{array}{ll}
            \Theta   & \mbox{if}  \quad U < p(\Theta) ,\quad \Theta<\pi\\
            \pi-\Theta & \mbox{if}\quad U > p(\Theta) ,\quad \Theta<\pi\\
            \Theta & \mbox{if} \quad U < p(\Theta) ,\quad \Theta>\pi\\
            3\pi-\Theta & \mbox{if} \quad U > p(\Theta) ,\quad \Theta>\pi
        \end{array}
    \right. $\\

 follows the CDF,  $H_2^*(y)=\dfrac{\left(y+\nu \sin{y}\right)}{2\pi},$ where $0<\nu\leq 1$, and $ 0<y<2\pi$.
\label{EAU_thm}	
\end{theorem}

\begin{proof}
    See the Appendix for the proof.
\end{proof}

\begin{figure}[b]
\centering
\subfloat[]{%
{\includegraphics[width=3.2 in]{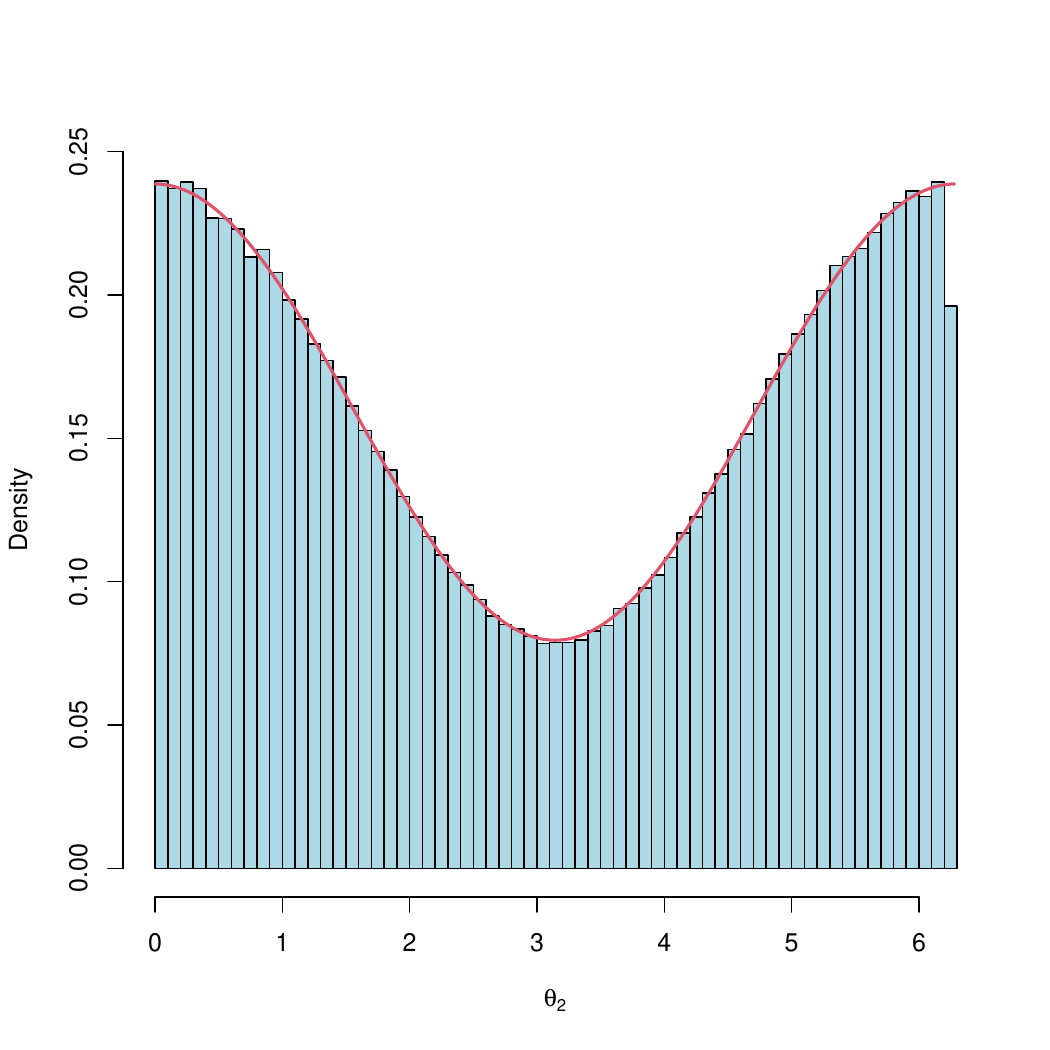 }}}
\subfloat[]{%
{\includegraphics[width=3.2 in]{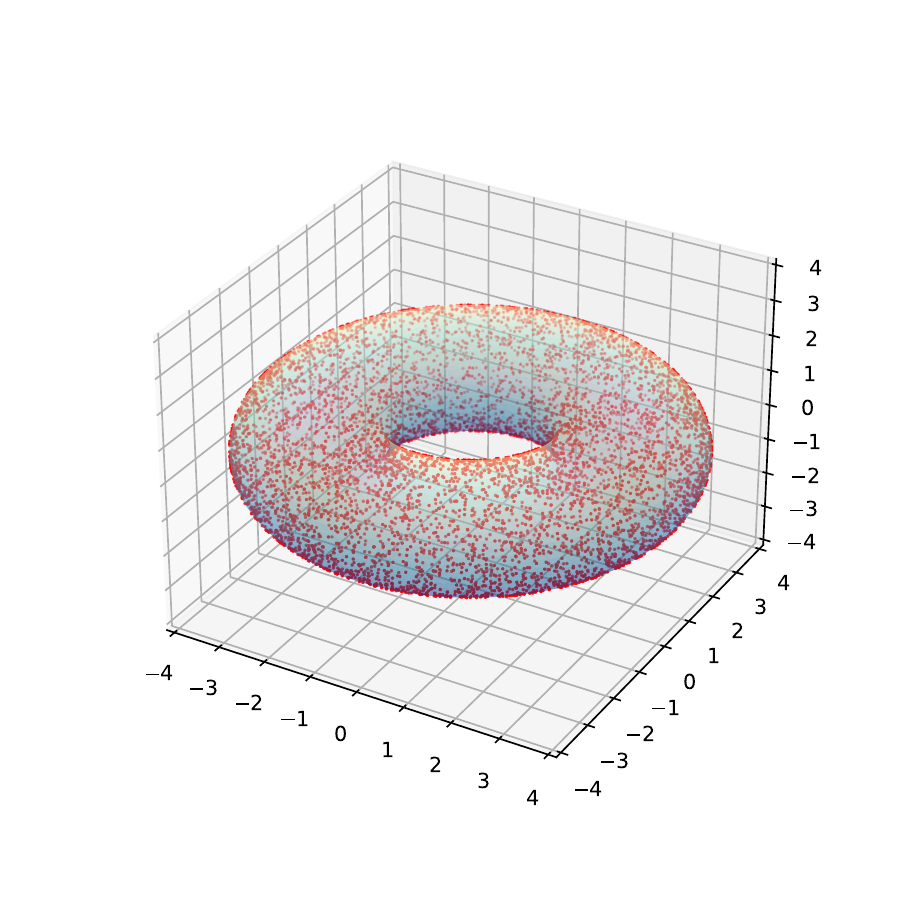}}}
\caption{ (a) Histogram of the sample from the marginal density $\frac{1}{2\pi}\left( 1+0.5~\cos{\theta} \right)$ 
 using Algorithm-\ref{algo_eau}.(b) The scatter plot of the uniformly distributed data on the surface of torus drawn using Algorithm-\ref{algo_eau}.} 
 \label{uni scatter}
\end{figure}

% \begin{table}[b]
% \caption{Acceptance percentage AUR sampling scheme by Diaconis et al \cite{diaconis2013sampling} from the curved torus, where the proposed exact sampling scheme has a 100\% acceptance rate.}\vspace{.5cm}
% \label{table:eau uni}
% \centering
% \begin{tabular}{lc|lr}
% \hline
% \textbf{a} && \textbf{AUR} &\\
% \hline
% $0.1$ && $50.35$ &\\
% $0.2$ && $51.07$ &\\
% $0.3$ && $49.91$ &\\
% $0.4$ && $49.68$ &\\
% $0.5$ && $50.14$ &\\
% $0.6$ && $49.82$ &\\
% $0.7$ && $49.88$ &\\
% $0.8$ && $49.41$ &\\
% $0.9$ && $49.96$ &\\
% $1$ && $49.41$ &\\
% \hline
% \end{tabular}
% \end{table}
Figure-\ref{uni scatter}(a) is the histogram of the sampled data from the marginal distribution of the vertical angle $\theta$ from the Equation- \ref{Cardioid_pdf}. Figure-\ref{uni scatter}(b) is the scattered plot of the data generated from the Algorithm-\ref{algo_eau} for $\theta$ and $\phi$ from the uniform distribution on $[0,2\pi]$,  maintaining the uniform distribution using area measure on the surface of a curved torus with $R=3, r=1.5$, and hence $a=0.5.$

\begin{table}[h!]
\scalebox{1}{
\centering
\begin{tabular}{|l|c c c c c c c c c c |}
\hline
\textbf{$\nu=\frac{r}{R}$} & 0.1 & 0.2 & 0.3 & 0.4 & 0.5 & 0.6 & 0.7 & 0.8 & 0.9 & 1\\
\hline
\textbf{AUR (\%)} & $50.35$ & $51.07$ & $49.91$ & $49.68$ & $50.14$ & $49.82$ & $49.88$ & $49.41$ & $49.96$ & $49.41$  \\
\hline
\end{tabular}}
\vspace{.5cm}
\caption{Acceptance percentage AUR sampling scheme by   \cite{diaconis2013sampling} from the curved torus, where the proposed exact sampling scheme has a 100\% acceptance rate.}
\label{table:eau uni}
\end{table}

\begin{figure}[h!]
\centering
\subfloat[]{%
{\includegraphics[width=3.2 in]{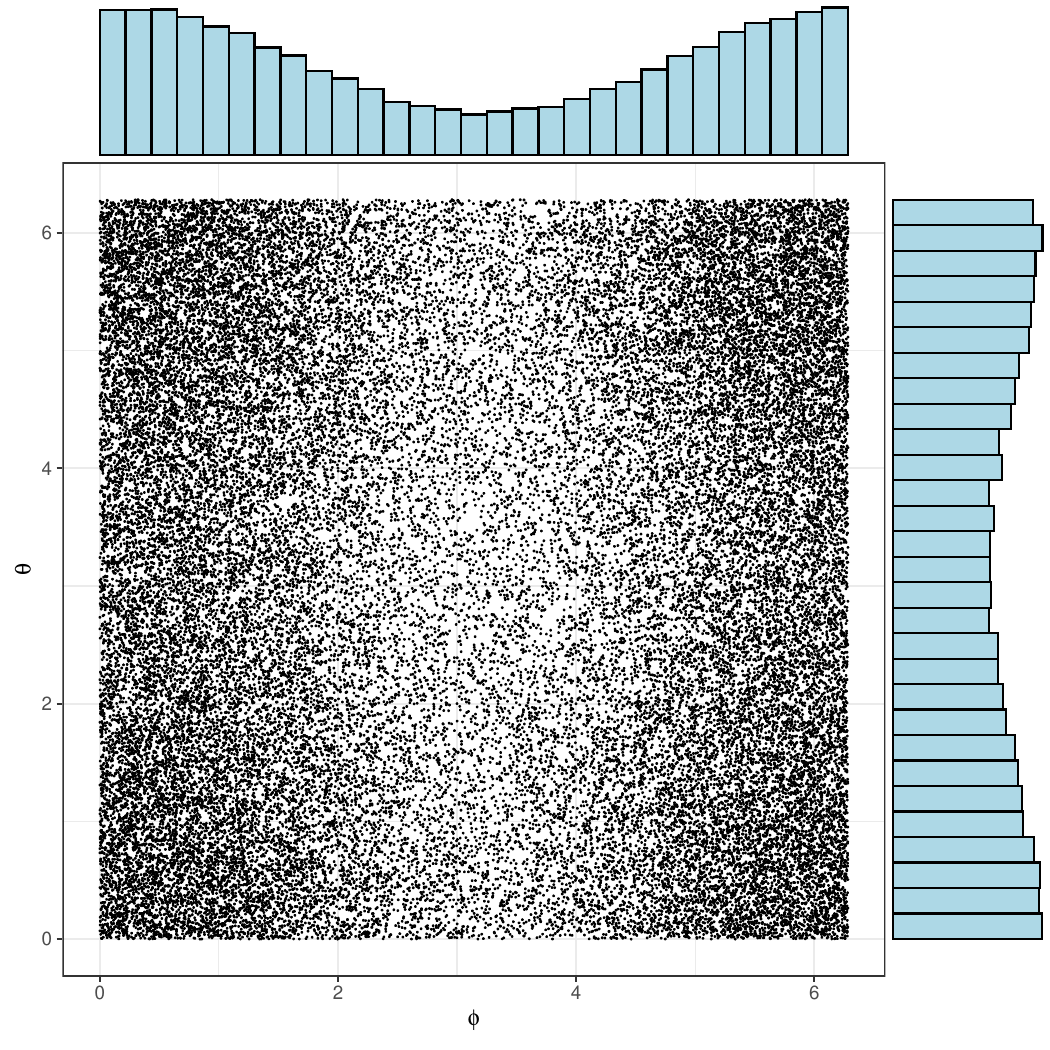}}}
\subfloat[]{%
{\includegraphics[trim= 80 80 80 80, clip,width=0.65\textwidth, height=0.45\textwidth]{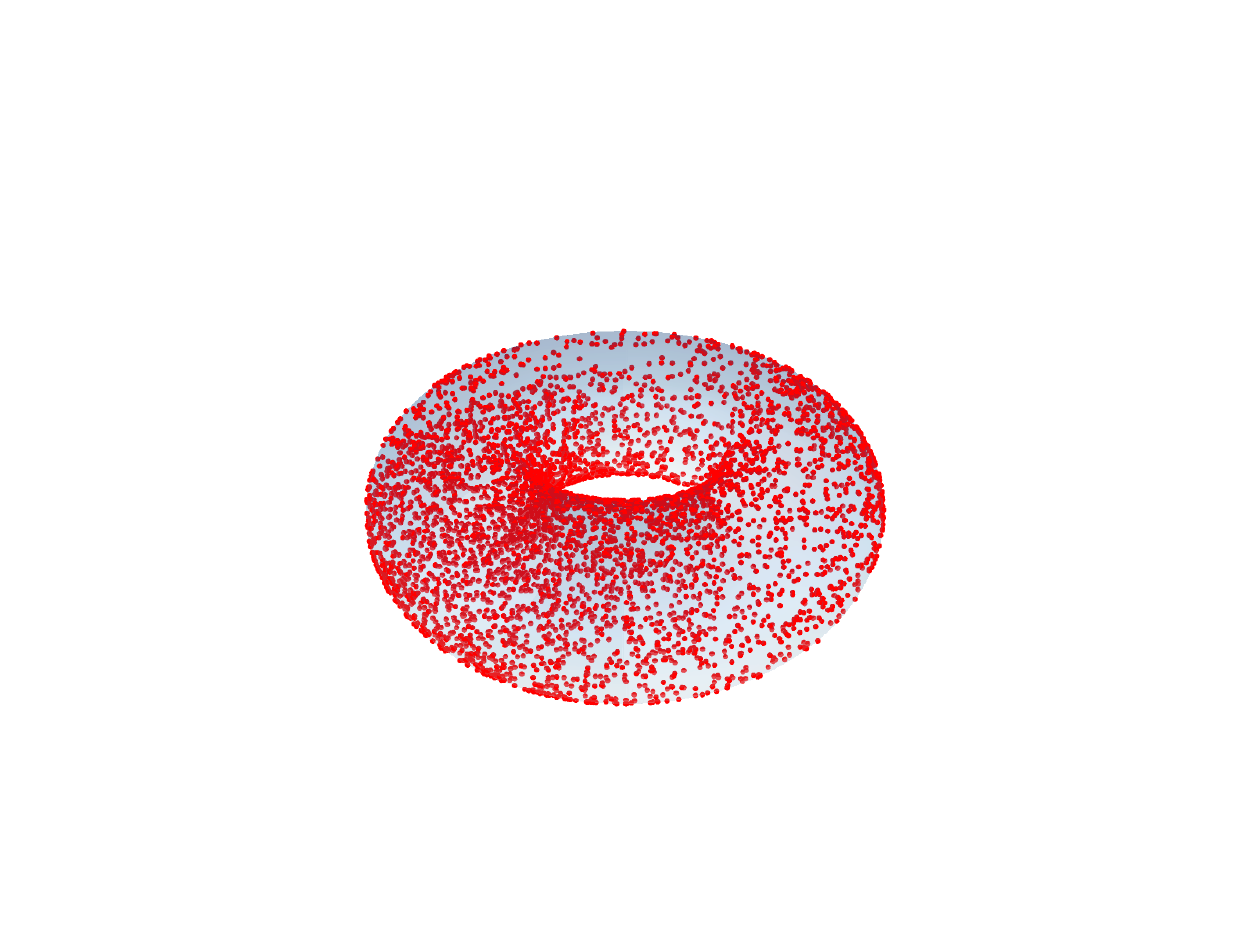}}}
\caption{ For the proposed bivariate density in Equation-\ref{dependent pdf} (a) shows the scatterplot and the histogram in the same figure, (b) displays the scatter plot for the  samples  on
the surface of a curved torus drawn using Algorithm-\ref{algo_eau}.} \label{bivariate_hist_scatter}
\end{figure}

\begin{rmk}
    Random sample $(\phi,\theta)$  from the joint density of the area-uniform distribution in Equation-\ref{area-uniform dist} can be drawn component wise from uniform $[0,2\pi]$ for $\phi$, and from $h_{2}^{*}(\theta)$ using Theorem-\ref{EAU_thm} for $\theta$. Similarly, we can generate from the proposed dependent model in Equation-\ref{dependent pdf} through proper conditioning and repeated uses of Theorem-\ref{EAU_thm} which is described in the following remark.
\end{rmk}
\begin{rmk}
    We use conditional distribution to get random samples from the proposed dependent model. Given that one of the marginals, namely the marginal for $\Theta$ in Equation-\ref{thet1 marginal cal} follows Cardioid distribution with a mean direction of \(\mu_1\), we begin by generation random samples from a Cardioid distribution with a mean direction of zero using the Algorithm-\ref{algo_eau}. Next, we add the mean direction \(\mu_1\) into the sample, therefore generating random samples from the density specified in Equation-\ref{thet1 marginal cal}. Now, we generate samples for the $\Phi$ using the conditional distribution of $\Phi$ given $\Theta=\theta$ as presented in Equation-\ref{phi conditional cal}. In this case, we will generate random samples from  Cardioid distribution with a zero mean direction.  Furthermore, we include the location $\frac{3\pi}{2}+\mu_2-\lambda(\theta-\mu_1)$ into the sample, where $\theta$ corresponds to the marginal density specified in Equation-\ref{thet1 marginal cal}. Finally, we obtain the samples $(\phi_i,\theta_i)$ for $i=1, \cdots, n. $ from the bivariate dependent model suggested in Equation-\ref{dependent pdf}. Figure-\ref{bivariate_hist_scatter}(a) illustrates the histogram and scatter in the same plot whereas Figure-\ref{bivariate_hist_scatter}(b) represents a scatterplot of the samples from the proposed bivariate density with $\nu=0.2, \kappa=-0.85,\lambda=0.46, \mu_1=0$, and $\mu_2=0$.
\end{rmk}

\subsection{Evaluating the performance of the estimators of the toroidal and regression models} \label{section parameter estimation simulated}
We employed  the\textit{ Nelder-Mead} algorithm to estimate the parameters $\nu$, $\kappa$, and $\lambda$, $\mu_1$, and $\mu_2$.
It is well known that this algorithm can get stuck in local optima. To mitigate this issue, we considered several random starting points and selected the estimates $\hat{\nu}$, $\hat{\kappa}$, and $\hat{\lambda}$, $\hat{\mu}_1$, and $\hat{\mu}_2$ that yielded the minimum AIC and BIC values. Table-\ref{table:simulated Aic Bic table} shows the different parameter values (standard errors in parentheses), the estimated parameter values from the \textit{Nelder-Mead} algorithm, along with the AIC and BIC values for sample sizes of 50, 100, 500, and 1000. 

\begin{figure}[t]
\centering
\subfloat[]{%
  \includegraphics[width=3 in]{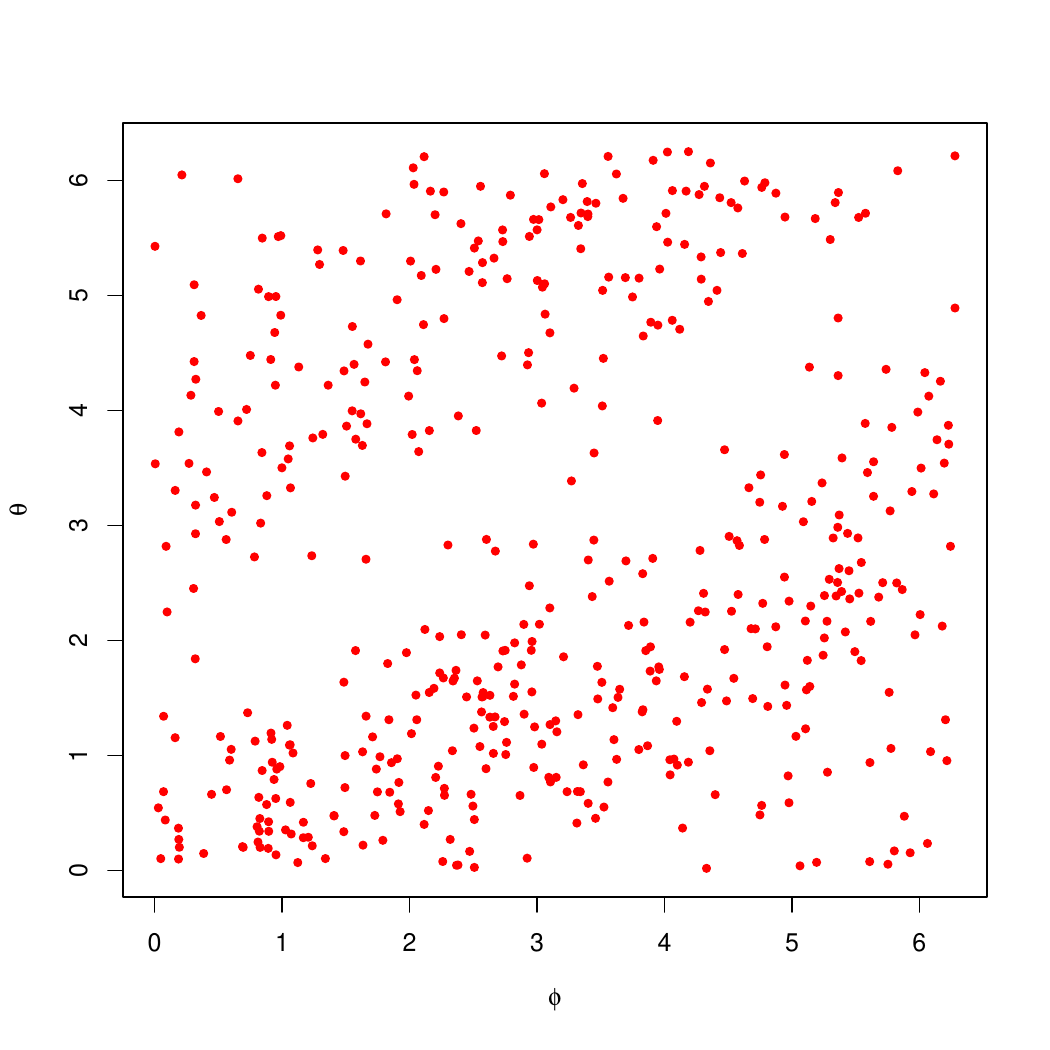}%
 
}
\subfloat[]{%
  \includegraphics[width=3 in]{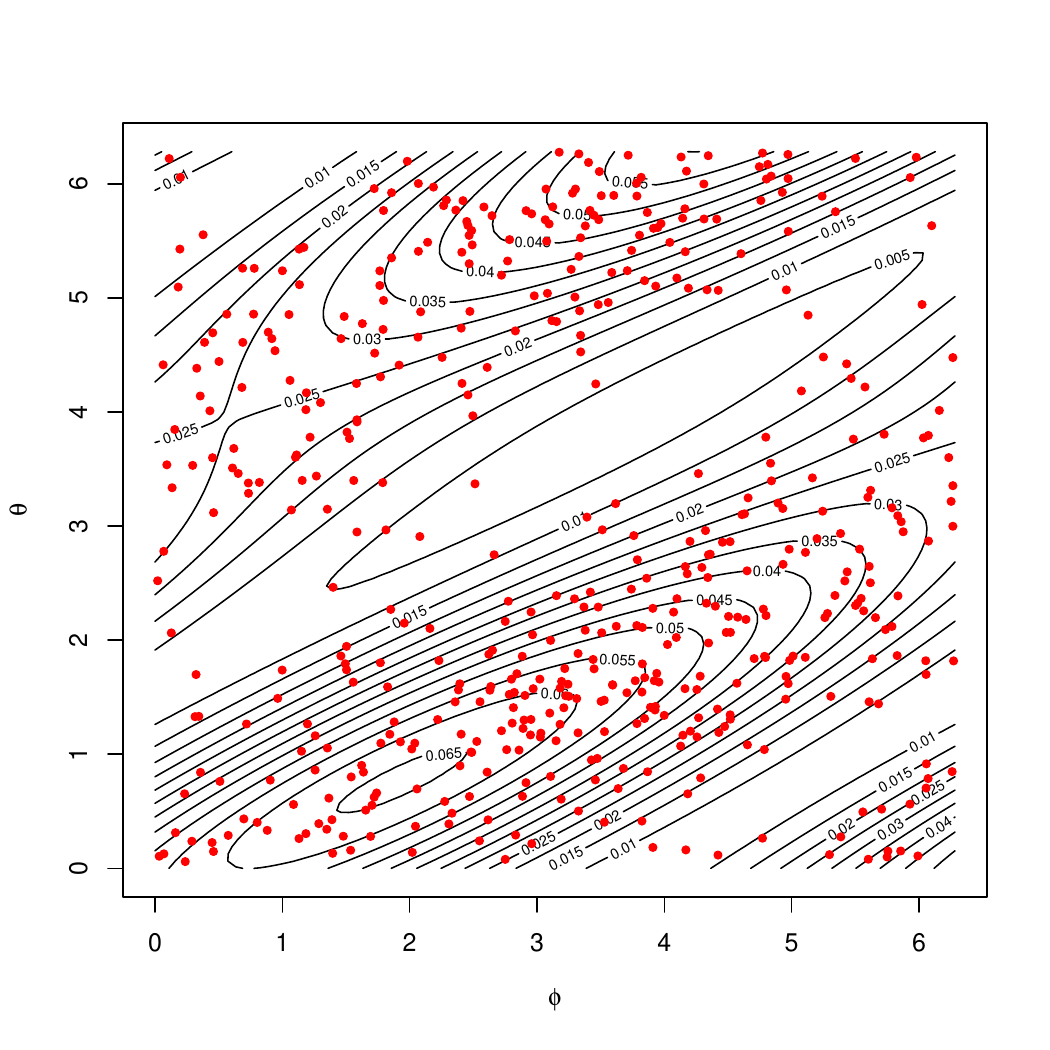}\hspace{2pt}
  
}\\
\subfloat[]{%
  \includegraphics[trim= 80 80 80 80, clip,width=0.6\textwidth, height=0.4\textwidth]{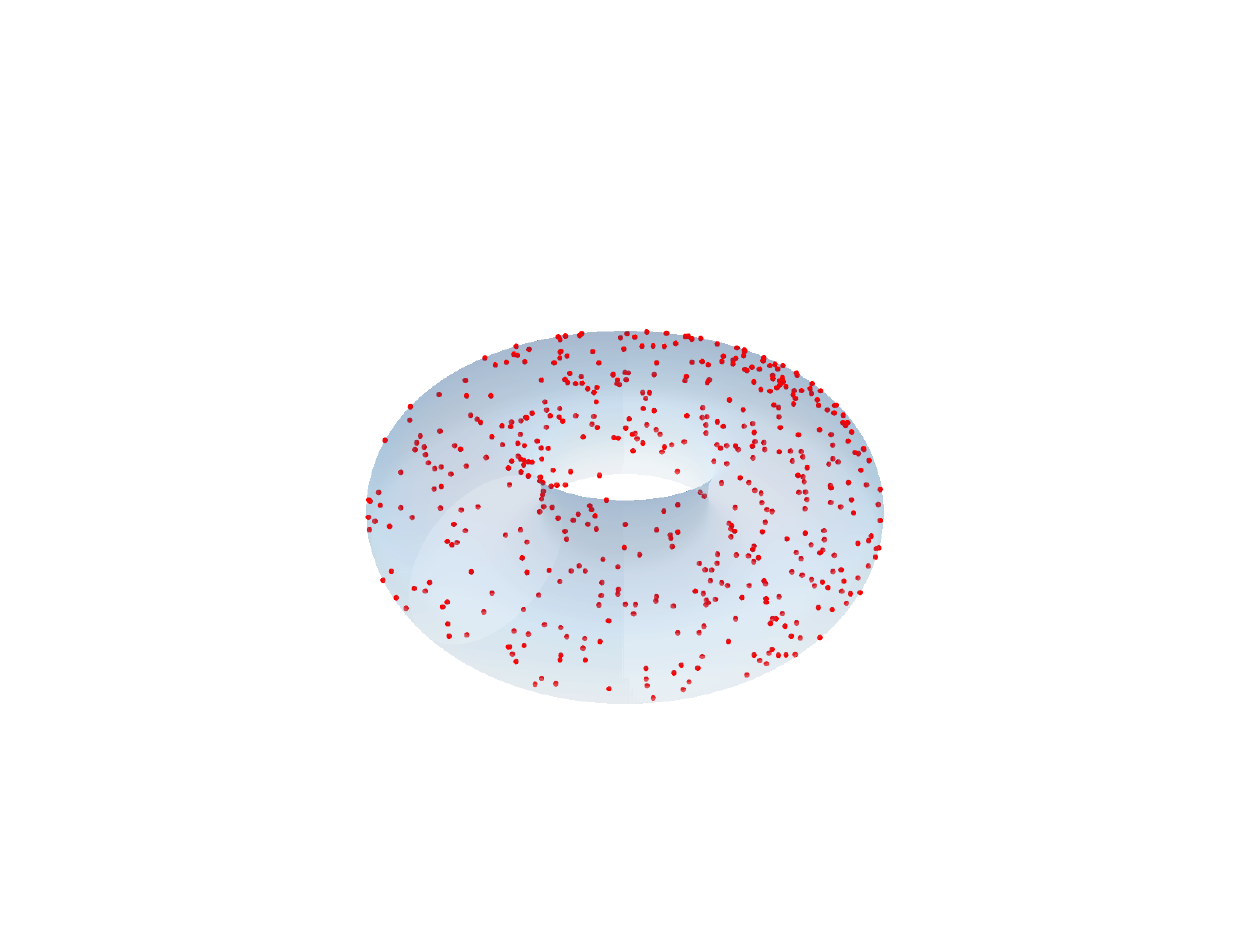}%
  
}
\subfloat[]{%
  \includegraphics[trim= 80 10 10 80, clip,width=0.6\textwidth, height=0.3\textwidth]{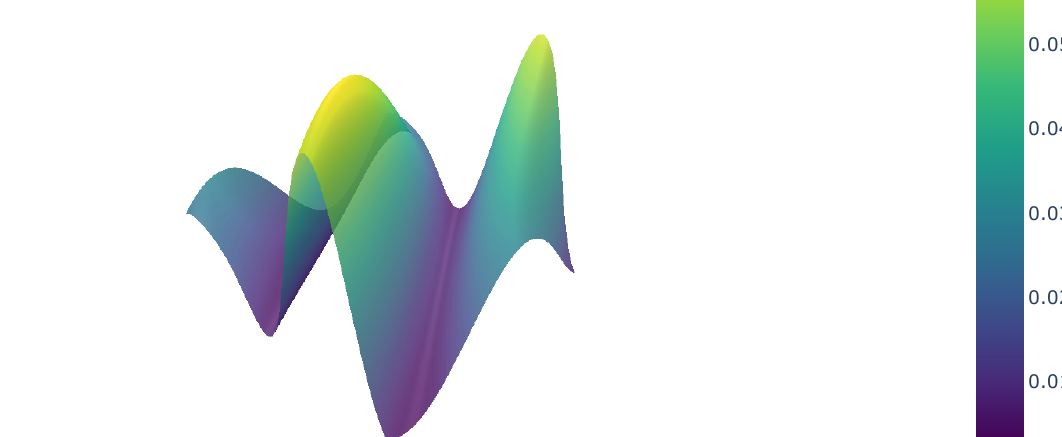}%
   
}

\caption{ (a) \& (c) represents the scatter plot of the simulated data on the flat torus and the curved torus, respectively. (b) \& (d) shows the contour and surface plot of the fitted density to the simulated data.}
\label{subfig:simulated_fit_figure}%
\end{figure}

Figure-\ref{subfig:simulated_fit_figure}(a) \& (c) represents the scatter plot of the simulated data from the proposed distribution on the flat torus and the curved torus for the sample size $n=500$ with the parameters $\nu=0.40,\kappa=0.80,
\lambda=-1.57$ and $\mu_1=0.80, \mu_2=3.6$ radians, respectively. Figure-\ref{subfig:simulated_fit_figure}(b) \& (d) shows the contour and surface plot of the fitted density with estimated values of the parameters as $\hat{\nu}=0.45, \hat{\kappa}=0.79, \hat{\lambda}=-1.56$, and  $\hat{\mu}_1=0.78, \hat{\mu}_2=3.51$ to the simulated data. Other figures can be obtained for the different sample sizes in a similar fashion.

\begin{figure}[t]
\centering
\subfloat[]{%
  \includegraphics[trim= 0 20 20 20, clip,width=0.4\textwidth, height=0.4\textwidth]{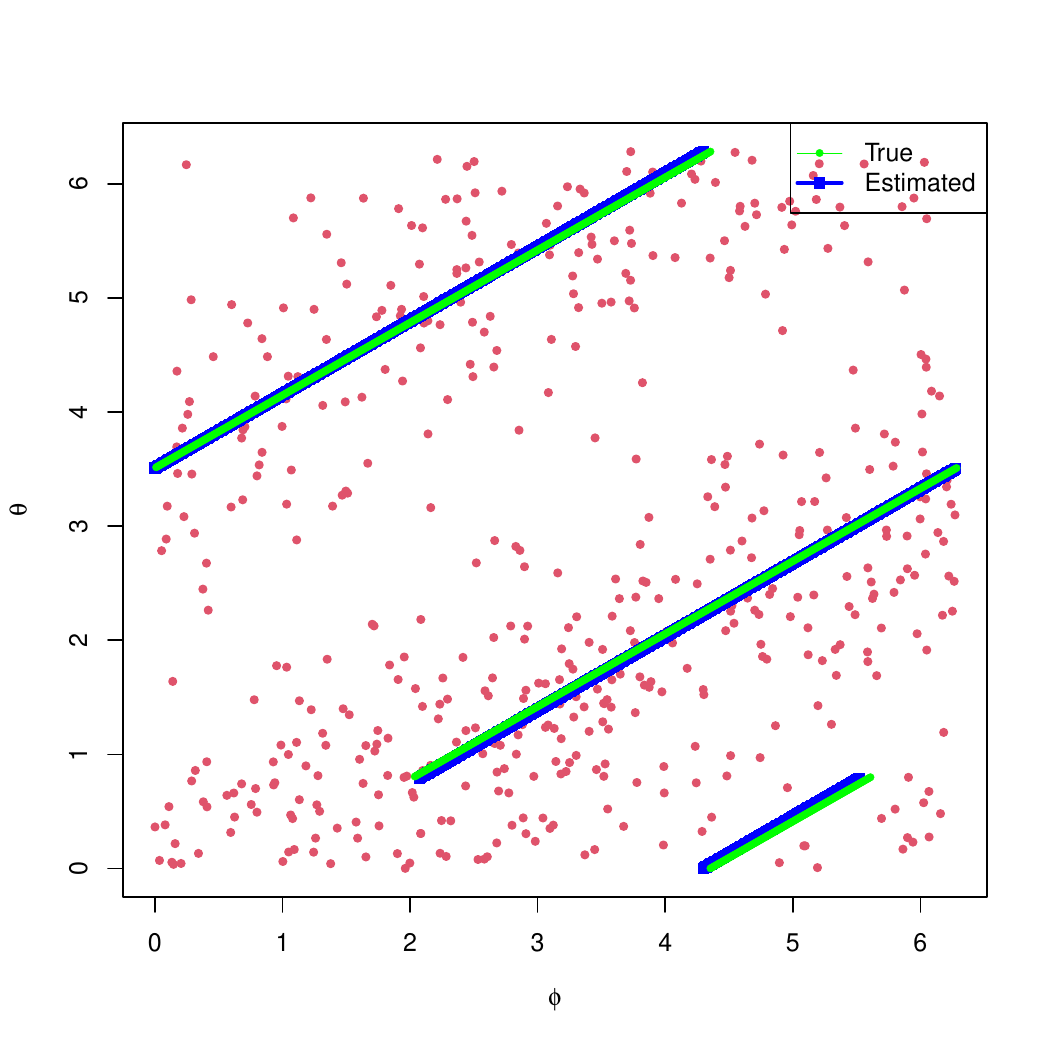}%
 
}
\subfloat[]{%
  \includegraphics[trim= 80 80 80 80, clip,width=0.65\textwidth, height=0.5\textwidth]{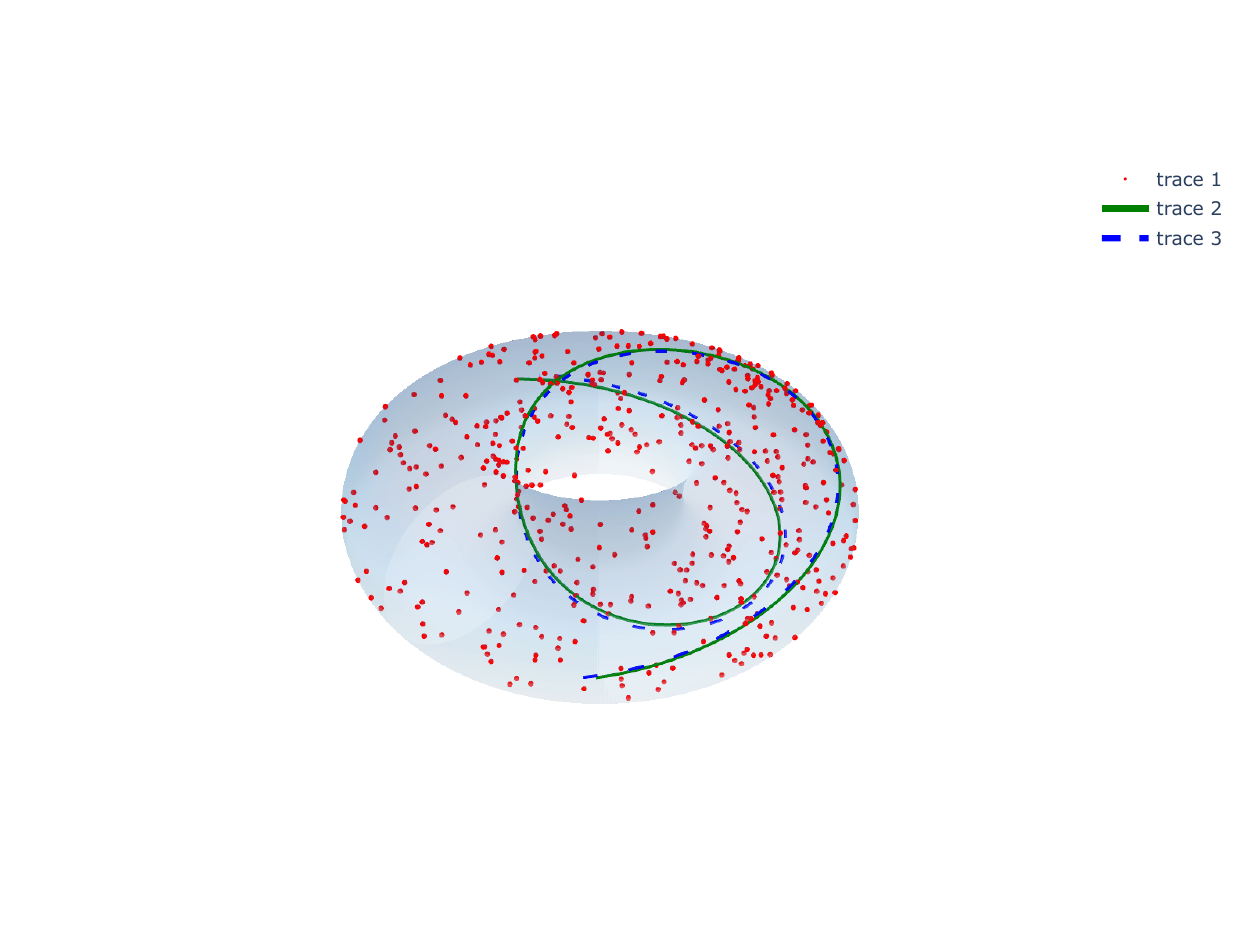}\hspace{2pt}
  
}
\caption{ (a) displays the plot of the exact (green line) and fitted ( blue line) regression curve on the simulated dataset on the flat torus, whereas (b) depicts the same on the curved torus.}
\label{subfig: regression curve simulated}%
\end{figure}

As we know, conditional expectation provides a framework for the regression model. Here, we assume that $\Phi$ is the response variable and $\Theta$
 represents the covariates, the regression model can be written as 
$E[\Phi |\Theta=\theta]=\mu_{\phi|\theta}=\frac{3\pi}{2}+\mu_2+\lambda(\theta-\mu_1)$, where $\theta$ follows the marginal density given in Equation-\ref{thet1 marginal cal}.
To obtain the regression curve, we use the estimated value $\hat{\lambda}$, and the samples $\theta_i$ for $i=1,\cdots, n$ are drawn from the marginal distribution of $\Theta$ as discussed in Section \ref{section random generate for bivariate}. Figure-\ref{subfig: regression curve simulated}(a) displays the plot of the exact (green line) and fitted (blue line) regression curve on the simulated dataset on the flat torus, whereas Figure-\ref{subfig: regression curve simulated}(b) depicts the same on the curved torus. From these figures, it is clear that both the regression curves are close to each other; hence, it indicates a good fit for this simulated dataset.

\begin{table}[t]
\centering
\resizebox{0.85\linewidth}{!}{%
  \begin{tabular}{|c|c|c|c|c|c|c|c|}
\hline
Sample size & Parameters value & Estimated value & Log-Likelihood & AIC & BIC  \\ \hline
\multicolumn{1}{ |c  }{\multirow{2}{*}{$n=50$} } &

\multicolumn{1}{ |c| }{$\nu=0.3$} & $\hat{\nu}=0.35~(0.21)$ & &  &   \\ 
\multicolumn{1}{ |c  }{}                        &
\multicolumn{1}{ |c| }{$\kappa=-0.4$}  & $\hat{\kappa}=-0.61~(0.17)$& &  &     \\ 
\multicolumn{1}{ |c  }{}                        &
\multicolumn{1}{ |c| }{$\lambda=1.3$} & $\hat{\lambda}=1.55~(0.14)$& -179.94 & 369.89 & 379.455 \\
\multicolumn{1}{ |c  }{}                        &
\multicolumn{1}{ |c| }{$\mu_1=0$} & $\hat{\mu}_1=0.36~(0.5)$& &  &     \\ 
\multicolumn{1}{ |c  }{}                        &
\multicolumn{1}{ |c| }{$\mu_2=0$} & $\hat{\mu}_2=0.29~(0.93)$& &  &     \\ 
\hline

\multicolumn{1}{ |c  }{\multirow{2}{*}{$n=100$} } &

\multicolumn{1}{ |c| }{$\nu=0.4$} & $\hat{\nu}=0.41~(0.13)$ & &  &    \\ 
\multicolumn{1}{ |c  }{}                        &
\multicolumn{1}{ |c| }{$\kappa=-0.60$}  & $\hat{\kappa}=-0.60~(0.11)$& &  &     \\ 
\multicolumn{1}{ |c  }{}                        &
\multicolumn{1}{ |c| }{$\lambda=-3.80$} & $\hat{\lambda}=-3.85~(0.12)$& -353.06& 716.11& 729.14 \\ 

\multicolumn{1}{ |c  }{}                        &
\multicolumn{1}{ |c| }{$\mu_1=0$} & $\hat{\mu}_1=6.28~(0.23)$& &  &     \\ 
\multicolumn{1}{ |c  }{}                        &
\multicolumn{1}{ |c| }{$\mu_2=4.25$} & $\hat{\mu}_2=4.16~(0.14)$& &  &     \\ 
\hline

\multicolumn{1}{ |c  }{\multirow{2}{*}{$n=500$} } &

\multicolumn{1}{ |c| }{$\nu=0.40$} & $\hat{\nu}=0.38~(0.06)$ & &  &    \\ 
\multicolumn{1}{ |c  }{}                        &
\multicolumn{1}{ |c| }{$\kappa=0.80$}  & $\hat{\kappa}= 0.81~(0.04)$& &    &   \\ 
\multicolumn{1}{ |c  }{}                        &
\multicolumn{1}{ |c| }{$\lambda=-1.57$} & $\hat{\lambda}=-1.55~(0.04)$& -1817.86& 3645.72& 3666.79   \\
\multicolumn{1}{ |c  }{}                        &
\multicolumn{1}{ |c| }{$\mu_1=0.80$} & $\hat{\mu}_1=0.79~(0.13)$& &  &     \\ 
\multicolumn{1}{ |c  }{}                        &
\multicolumn{1}{ |c| }{$\mu_2=3.60$} & $\hat{\mu}_2=3.64~(0.22)$& &  &     \\ 
\hline

\multicolumn{1}{ |c  }{\multirow{2}{*}{$n=1000$} } &

\multicolumn{1}{ |c| }{$\nu=0.8$} & $\hat{\nu}=0.78~(0.03)$ & &  &    \\ 
\multicolumn{1}{ |c  }{}                        &
\multicolumn{1}{ |c| }{$\kappa=0.7$}  & $\hat{\kappa}=0.73~(0.03)$& &     &   \\ 
\multicolumn{1}{ |c  }{}                        &
\multicolumn{1}{ |c| }{$\lambda=2.1$} & $\hat{\lambda}=2.12~(0.03)$& -3515.15& 7040.31& 7064.85   \\

\multicolumn{1}{ |c  }{}                        &
\multicolumn{1}{ |c| }{$\mu_1=1.5$} & $\hat{\mu}_1=1.52~(0.05)$& &  &     \\ 
\multicolumn{1}{ |c  }{}                        &
\multicolumn{1}{ |c| }{$\mu_2=1.5$} & $\hat{\mu}_2=1.53(0.09)$& &  &     \\ 
\hline

\end{tabular}}
\vspace{0.3cm}
\caption{Contains different parameter values and corresponding estimated values (standard errors in parentheses) of the parameters using the Nelder-Mead algorithm along with the Log-Likelihood, AIC, and BIC values for different sample sizes of $50, 100, 500,$ and $1000,$ respectively.}
\label{table:simulated Aic Bic table}
\end{table}

\section{Astigmatism data analysis} \label{section data analysis}
During cataract surgery, the incision made in the cornea can sometimes alter its natural shape, leading to astigmatism. This phenomenon was specifically observed when small incision cataract surgery (SICS) was performed, whether using the Vectis or Snare technique. In contrast to the above techniques, this issue did not occur when the surgery was conducted using Conventional or Torsional Phacoemulsification techniques. 
To assess this surgically induced astigmatism (SIA) during the follow-up period, SIA software   was utilized. As a result, our data analysis will focus exclusively on the data obtained from SICS procedures using the Vectis and Snare techniques.

The cataract surgery dataset we have considered for this analysis obtained from both procedures consists of 40 observations of a single measurement of the axis of astigmatism before surgery, followed by a measurement taken one month and three months after surgery. The dataset originally included three missing values: one patient did not attend the follow-up, and two patients only attended the one-month post-surgery follow-up. As a result, we discarded the data for the patients who did not attend any follow-ups. For the two patients who missed the three-month follow-up, we imputed the missing values with the circular mean of the rest of the measurements taken three months after surgery. Consequently, the dataset now includes a total of 39 observations, allowing for a detailed analysis of the progression of astigmatism over time.

\subsection{Toroidal model fitting} \label{section model fitting}
\begin{figure}[t]
\centering
\subfloat[]{%
  \includegraphics[width=2.8 in]{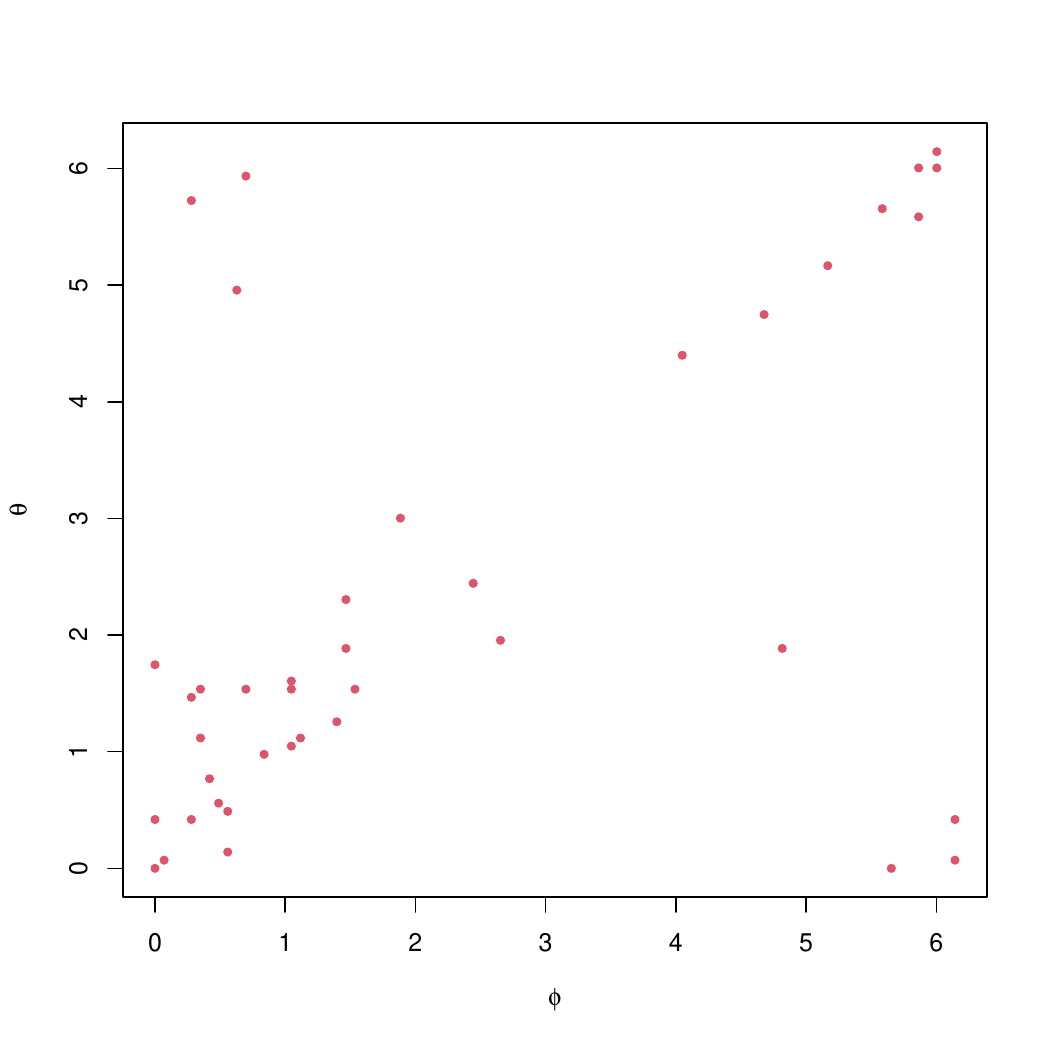}%
 
}
\subfloat[]{%
  \includegraphics[width=2.8 in]{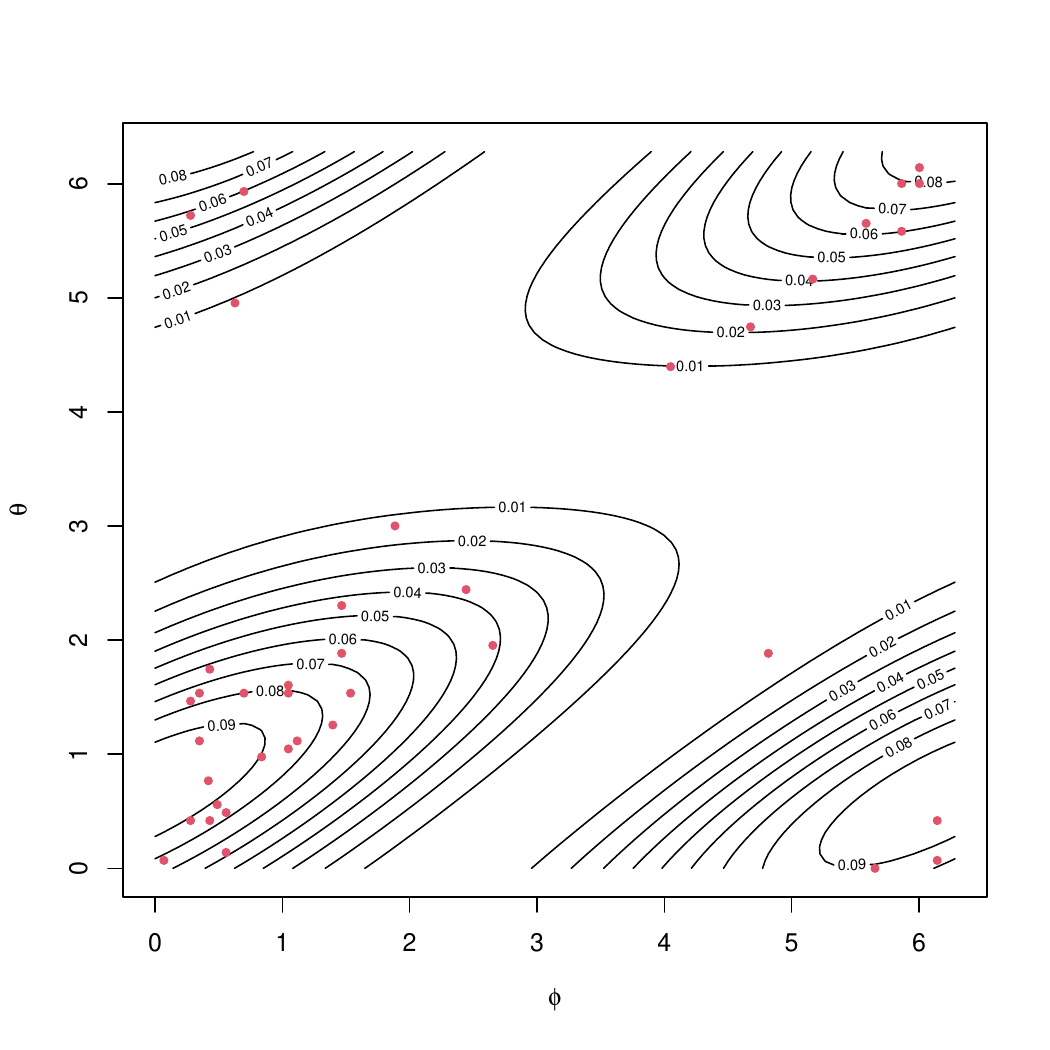}\hspace{2pt}
  
}\\

\caption{ (a)  display the planar plot of the measurements of the axis of astigmatism taken after the first month of surgery and three months after surgery. (b) show the contour plot of the proposed bivariate density given in Equation-\ref{dependent pdf}, fitted to the dataset.}
\label{subfig:model fitting figure}%
\end{figure}

In this section, we fit the proposed bivariate probability density function to the dataset. Let $\phi_i$ for $i=1,\cdots, 39$ represent the measurements of the axis of astigmatism taken three months after surgery, and let $\theta_i$ for $i=1,\cdots, 39$ represent the same measurements taken one month after surgery.
 This data is used in the \textit{Nelder-Mead} algorithm to estimate the parameters $\nu$, $\kappa$, $\lambda$, $\mu_1$, and $\mu_2$. Since the algorithm can get stuck in local optima, we considered several random initial points and selected the estimated values of the parameters $\hat{\nu}$, $\hat{\kappa}$, $\hat{\lambda}$, $\hat{\mu}_1$ and $\hat{\mu}_2$ that yield the lowest AIC and BIC values.
The Table-\ref{table:bivariate_fit_data} presents the maximum likelihood estimates (MLE) of the parameters, the maximized log-likelihood (log L), Akaike Information Criterion (AIC), and Bayesian Information Criterion (BIC) values for both datasets.
Figure-\ref{subfig:model fitting figure}(a) represents the planar plot of the astigmatism dataset. This plot demonstrates the dependence between the measurements of the axis of astigmatism taken after the first month of surgery and after the third month of surgery for the dataset. Figure-\ref{subfig:model fitting figure}(b) depicts the fitted proposed bivariate density, which appears to show a satisfactory fit to the dataset. The relatively large value of the parameter $\lambda$, which controls the dependence between the two circular variables, indicates a strong association between the measurements of the axis of astigmatism taken one month and three months after surgery. This is further supported by the circular correlation coefficient between these two-time points, calculated as $0.6979$, reinforcing the conclusion that the measurements are indeed strongly correlated.

\begin{table}[h!]
\centering
\scalebox{0.75}{
\begin{tabular}{|l|c c c c c c c c|}
\hline
\textbf{Estimated parameters} & $\hat{\nu}$ & $\hat{\kappa}$ & $\hat{\lambda}$ & $\hat{\mu}_1$ & $\hat{\mu}_2$ &$ \log L$ & AIC & BIC\\
\hline
\hline
\textbf{Values} & $0.98~(0.19)$ & $0.98~(0.13)$ & $-1.14~(0.09)$ & $0.64~(0.17)$ & $1.47~(0.44)$ & $-112.89$ & $ 235.79 $ & $244.11$ \\
\hline
% \textbf{Dataset-II (Measurement-II)} & $0.78$ & $1$  & $-1.37$ & $0.74$ & $0.62$ & $23.29$ & $-36.60$ & $-28.54$ \\
% \hline
\end{tabular}}
\vspace{.1cm}
\caption{Maximum likelihood estimates (MLE) of the parameters (standard errors in parentheses), the maximized log-likelihood (Log L), Akaike information criterion (AIC), and Bayes information criterion (BIC) for the proposed model.}
\label{table:bivariate_fit_data}
\end{table}

% Average perpendicular distance for proposed= 33.14561

\subsection{Regression model on the surface of torus}
\begin{figure}[t]
\centering
\subfloat[]{%
  \includegraphics[width=2.8 in]{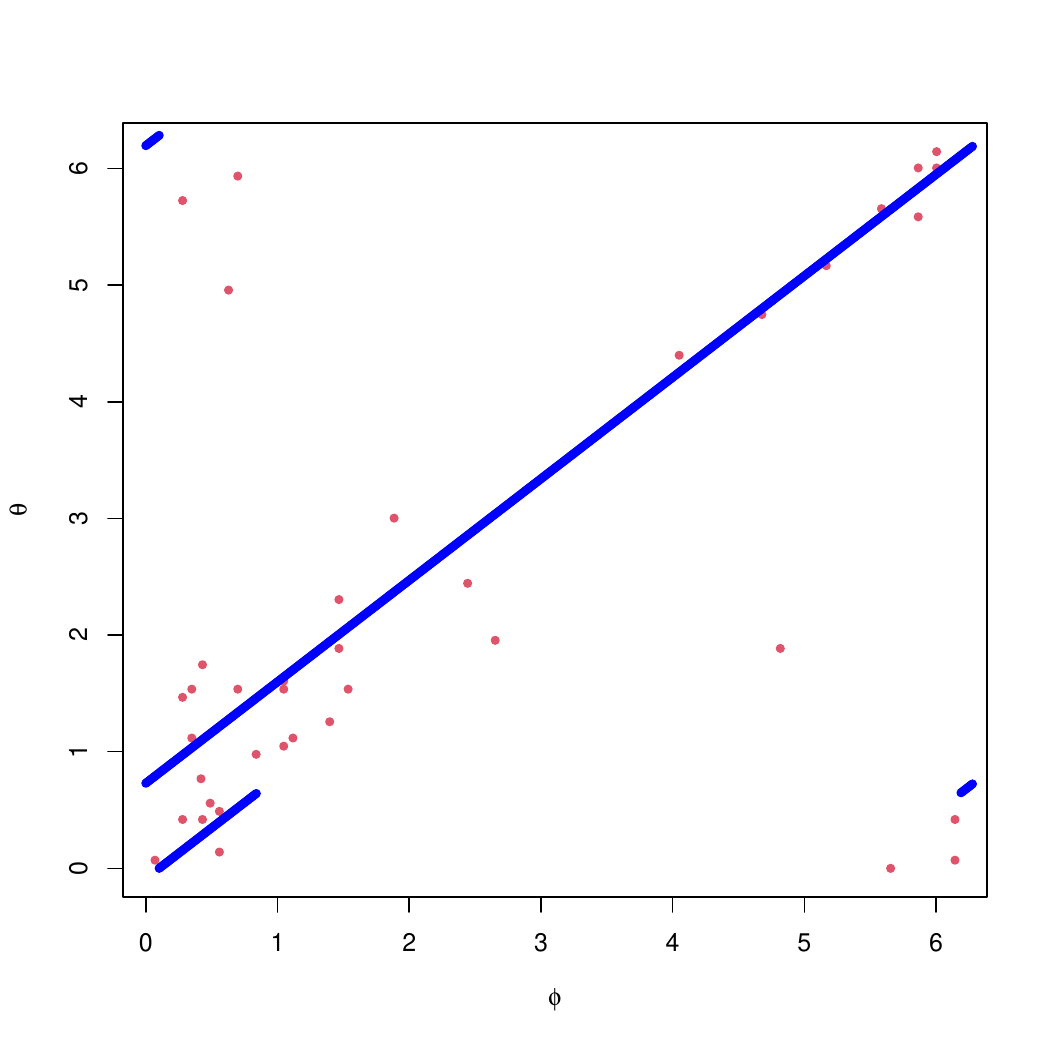}
  
}
\subfloat[]{%
  \includegraphics[trim= 90 90 90 90, clip,width=0.6\textwidth, height=0.5\textwidth]{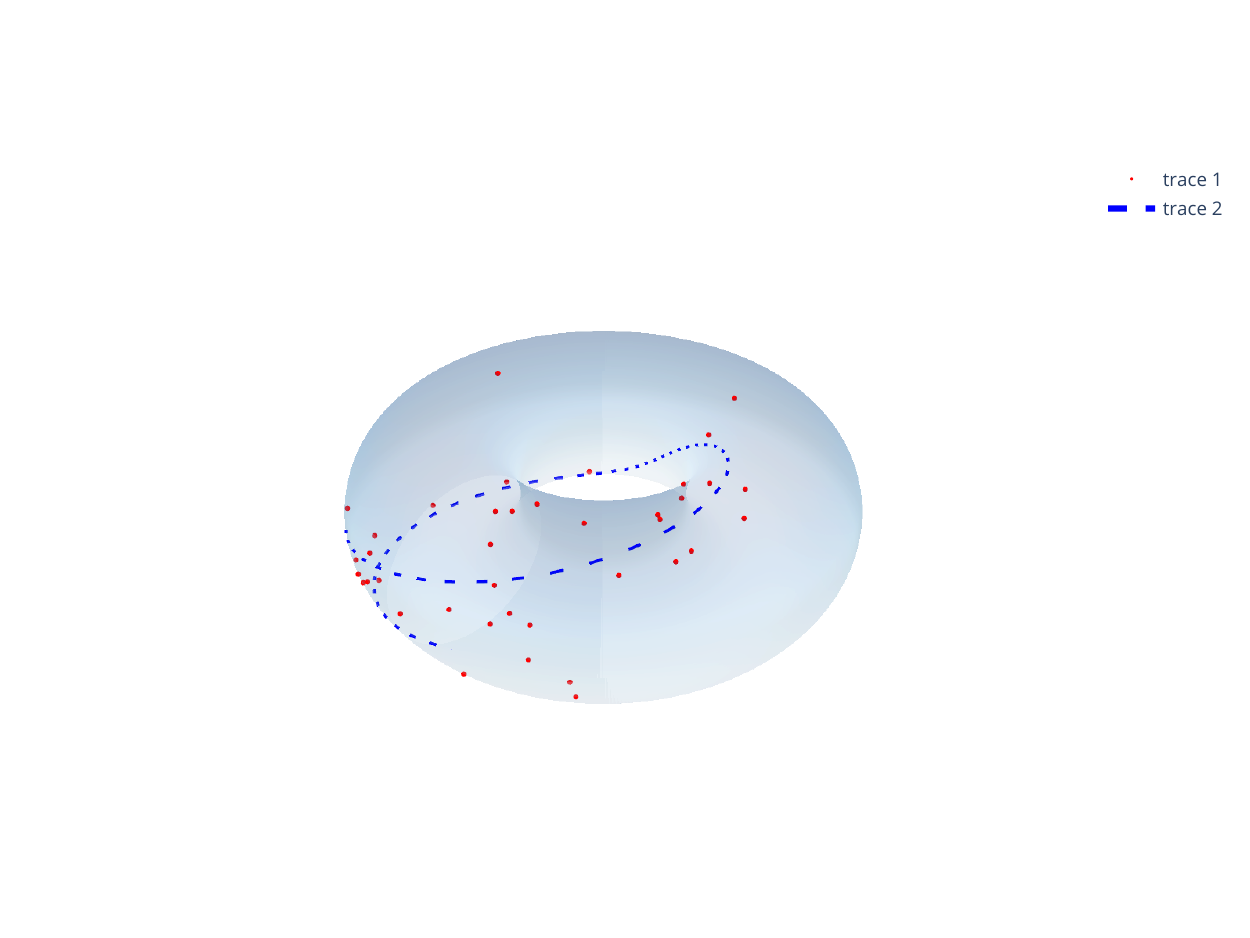}\hspace{2pt}
  
}

\subfloat[]{%
  \includegraphics[width=2.8 in]{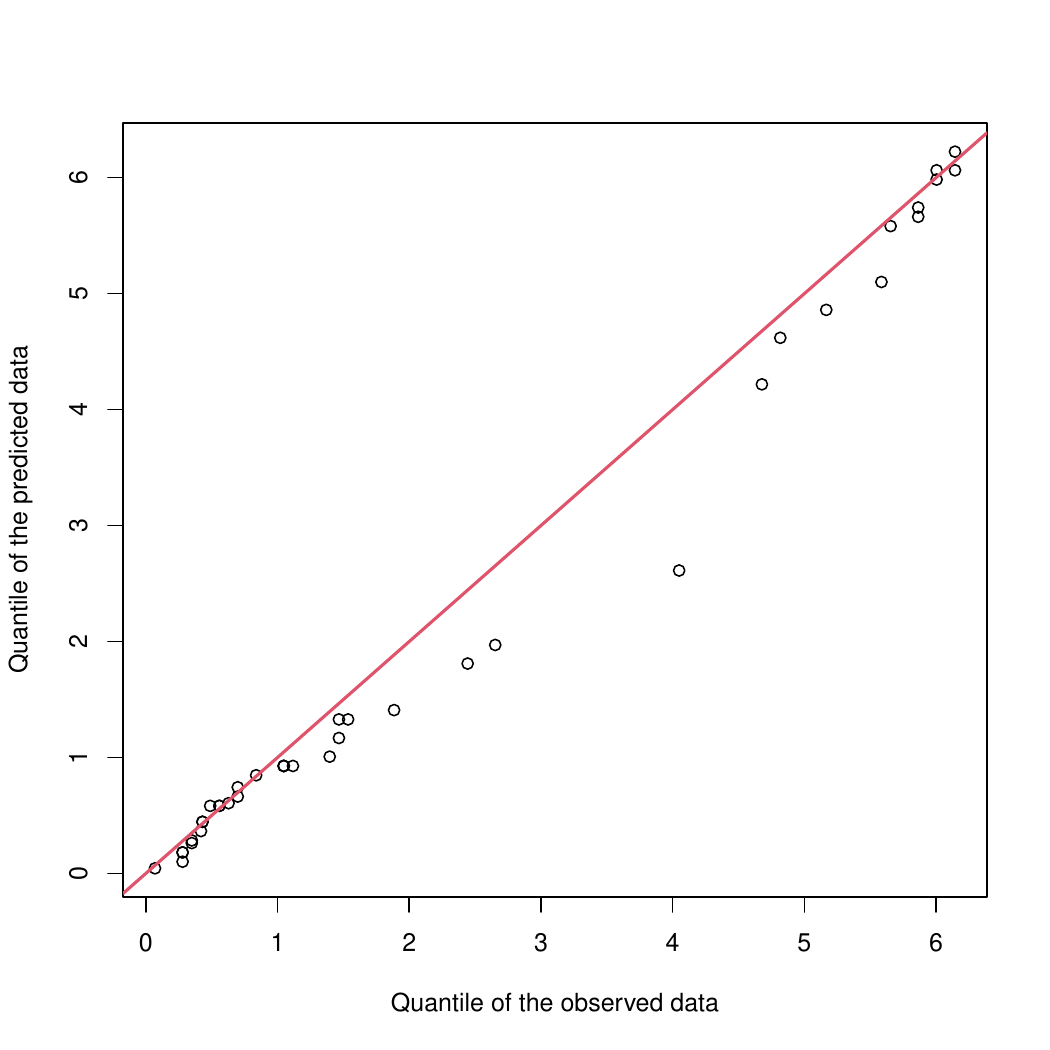}
  
}
\subfloat[]{%
  \includegraphics[width=2.8 in]{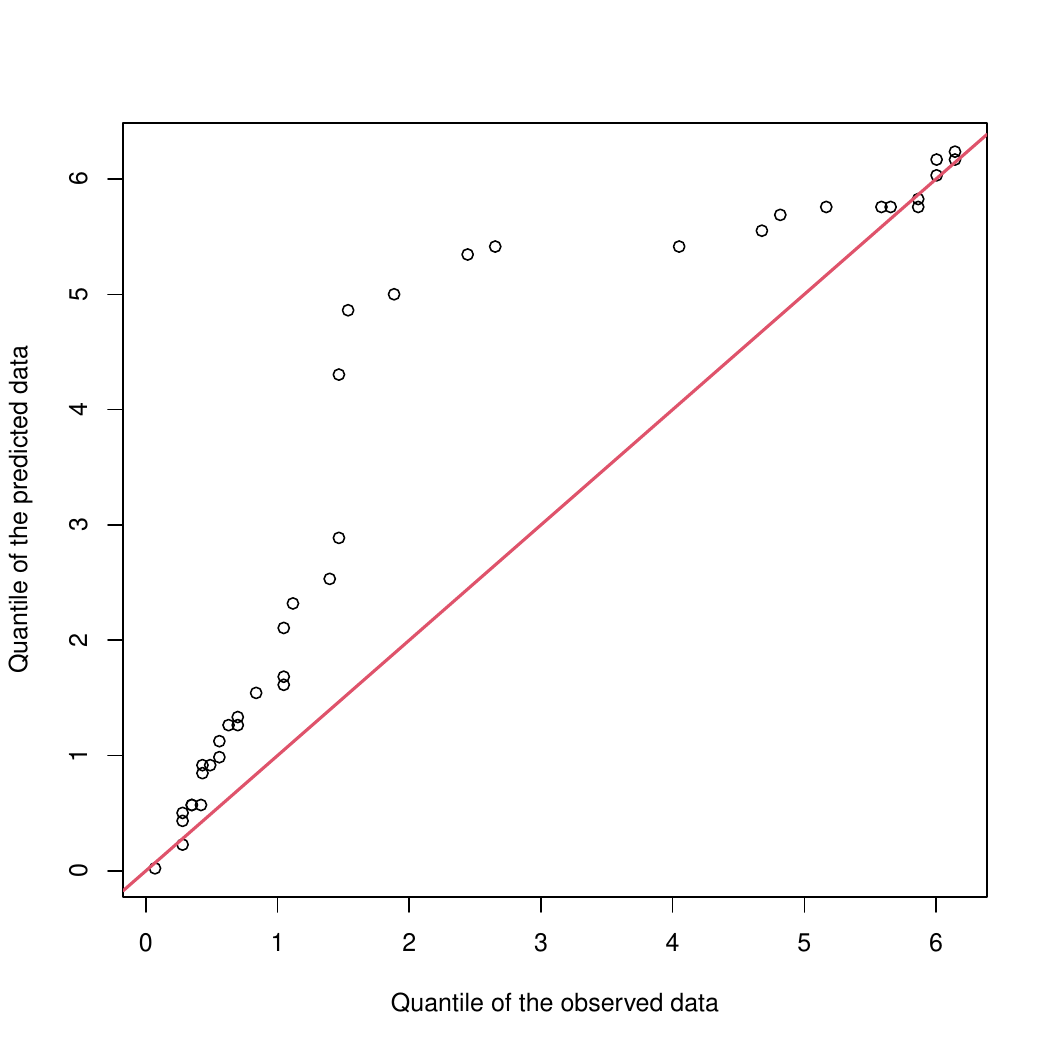}
  
}
\caption{ (a) displays the plot of the fitted (solid blue line) regression curve on the astigmatism dataset on the flat torus, whereas (b) depicts the same on the curved torus. (c) \& (d) represents the QQ plots of 39 points in the radian scale of observed data versus predicted data by the proposed model and the model by \cite{kato2008circular}, respectively.}
\label{subfig: regression model data figure}%
\end{figure}

For this set of data, the regression model provides the conditional mean direction of the circular random variable $\Phi$ (response), representing the astigmatism axis three months after surgery, given $\Theta=\theta$ (covariate), representing the astigmatism axis one month after surgery.
Since the conditional expectation depends only on the mean directions $\mu_1$, $\mu_2$, and the dependence parameter $\lambda$, we will use the estimated values of $\hat{\mu}_1$, $\hat{\mu}_2$, and $\hat{\lambda}$ from Section \ref{section model fitting}. Therefore, the regression model for the dataset is
$$\hat{\mu}_{\phi|\theta}=\frac{3\pi}{2}+1.47-1.14~(\theta- 0.64).$$
Figure-\ref{subfig: regression model data figure}(a) displays the plot of the fitted (blue line) regression curve on the astigmatism dataset on the flat torus, whereas Figure-\ref{subfig: regression model data figure}(b) depicts the same on the curved torus.

One of the well-known circular-circular regression models is due to  \cite{kato2008circular} utilizing the Mobius transformation and using wrapped Cauchy distribution because of its elegant properties. We have compared our proposed regression model with the aforementioned one by QQ-plot.  In the   Figure-\ref{subfig: regression model data figure}(b), we present the QQ-plot between the observed $\phi$-values and the predicted $\phi$-values generated using the proposed model; while, in the   Figure-\ref{subfig: regression model data figure}(c), 
we present the QQ-plot between the observed $\phi$ values and the predicted $\phi$ values generated using the model by   \cite{kato2008circular}.
 In the QQ plots, the quantiles of the observed data are along the horizontal axis, while the quantiles of the predicted values are along the vertical axis. To quantify the accuracy of the fitted regression model through QQ-plot, we consider the measure of the average perpendicular distance of the plotted data, i.e. (quantile of the observed data, quantile of the predicted data), to the line with radiant $1$ and passing through the origin. We observe that the average value obtained for the proposed model is 33.2,  whereas that of the model by Kato is 82. Clearly, the proposed model gives a better fit to the data. 
 The proposed regression model achieves an AIC of $113.275$ and a BIC of $119.930$, substantially outperforming the model proposed by  \cite{kato2008circular}, which has an AIC of $148.970$ and a BIC of $155.624$.
 These metrics decisively highlight the better fit of our approach to this dataset.

\begin{rmk}
It is worth noting that although the theoretical expression for the conditional expectation or mean direction of  $\Theta$ given $\Phi=\phi$ is known, as shown in Theorem-\ref{conditional of theta thm}, we do not use it for the regression model. This is because the measurements of the axes of astigmatism one month and three months after the surgery are not ordered. Consequently, we could use $\Phi$ (response) as the axis of astigmatism after the first month of surgery and $\Theta$ (covariate) as the astigmatism axis after the third month of surgery.
\end{rmk}

\section{Conclusion}
\label{conclusion}
We have initiated our study by revisiting the area-uniform toroidal distribution on the surface of a curved torus and emphasizing the natural importance of including area elements in the distribution. By transforming the area-uniform distribution, we have developed a five-parameter toroidal distribution that uses its intrinsic geometry to model the distribution of two dependent circular random variables on the torus. We have shown that for the proposed model, the marginal distributions are Cardioid and one of the conditional random variables follows the Cardioid distribution.  This property enabled us to propose a circular-circular regression model based on conditional expectation. We introduced an exact sampling method to draw random samples from the proposed marginal densities of Cardioid distribution using a probabilistic transformation which is different from the inverse cumulative distribution function transformation. It gives a huge computational advantage over the existing acceptance-rejection sampling scheme with a high rejection rate, approximately $50\%$, for the  Cardioid distribution.  Furthermore, we successfully generated random samples from the proposed toroidal distribution using the conditional distribution.  
 Finally, for the practical applicability of the model, the proposed bivariate distribution and the regression model have been implemented on astigmatism data resulting from cataract surgery. This work not only addresses specific challenges in medical data analysis but also contributes to the broader field of circular statistics by offering a potential model for bivariate dependent circular variables as well as a conditional distribution-based regression model.

\section{Acknowledgments}

The author, S. Biswas, acknowledges and appreciates the financial assistance provided in the form of a junior/senior research fellowship by the Ministry of Human Resource and Development (MHRD) and IIT Kharagpur, India.
%%%%%%%%%%%%%%%%%%%%%%%%%%%%%%%%%%%%%%%%%%%%%%
%% Funding information, if any,             %%
%% should be provided in the                %%
%% funding section.                         %%
%%%%%%%%%%%%%%%%%%%%%%%%%%%%%%%%%%%%%%%%%%%%%%
\section{Funding}

 Author B. Banerjee would like to thank the Science and Engineering Research Board (SERB), Department of Science \& Technology, Government of India, for the MATRICS grant (File number MTR/2021/000397)  for the project's funding.

% %\backmatter
% \bmsection*{Author contributions}

% This is an author contribution text. This is an author contribution text. This is an author contribution text. This is an author contribution text. This is an author contribution text.

\section*{Conflict of interest}

The authors declare no potential conflict of interest.

\begin{scriptsize}
\bibliographystyle{abbrvnat}
	\bibliography{buddha_bib.bib}
\end{scriptsize}

\section{Appendix}
 \textbf{Intrinsic geometry of torus}
\label{appendix}

The parametric equation of $2$-dimensional curved torus  is the Lipschitz image \cite[see][]{diaconis2013sampling} of the set $\{ 
 (\phi,\theta):0<\phi,\theta<2\pi\}\subset \mathbb{R}^2.$ Clearly, the function $f(\phi,\theta)=\{  (R+r\cos{\theta})\cos{\phi}, (R+r\cos{\theta})\sin{\phi}, r\sin{\theta} \}$ is a differentiable function from $\mathbb{R}^2$ to $\mathbb{R}^3$. Now, the partial derivatives of $f$ with respect to $\phi$, and $\theta$ are 
$$ \dfrac{\partial f}{\partial \phi}=\{ -(R+r\cos{\theta})\sin{\phi}, (R+r\cos{\theta})\cos{\phi}, 0 \}, $$ and
$$\dfrac{\partial f}{\partial \theta}=\{  -r\sin{\theta} \cos{\phi}, -r\sin{\theta}\sin{\phi}, r\cos{\theta}  \},$$ respectively. Hence, the derivative matrix is

$$Df(\phi,\theta)=\begin{bmatrix}
 -(R+r\cos{\theta})\sin{\phi} &  -r\sin{\theta} \cos{\phi} \\
 (R+r\cos{\theta})\cos{\phi} & -r\sin{\theta}\sin{\phi}\\
  0& r\cos{\theta}
\end{bmatrix}.$$ 

Therefore, the square of the Jacobian can be calculated as

\begin{eqnarray}
       J_{2}^{2}f(\phi,\theta)&=& \text{det} \left[ D^Tf(\phi,\theta) \cdot Df(\phi,\theta)\right]\nonumber\\
       &=& \text{det} \begin{bmatrix}
 (R+r\cos{\theta})^{2} & 0 \\
 0 & r^{2}
\end{bmatrix} \nonumber\\
&=&r^{2}(R+r\cos{\theta})^{2}
\end{eqnarray}

Using the above expression, we get the area element as 

\begin{equation}
    dA=r(R+r\cos{\theta}) d\phi d\theta,
    \label{area_element}
\end{equation}
 which is the square root of the determinant of the product of transpose of the derivative matrix and the derivative matrix itself. 
 \cite{diaconis2013sampling} proposed to draw the samples  $(\phi,\theta)$ from the density function given  in Equation- \ref{decompose equation} to ensure the uniformity with respect to area measure on the surface of a curved torus
\begin{equation}
    h^*(\phi,\theta)=\dfrac{(1+(r/R) \cos{\theta})}{4\pi^{2}}=g_1(\phi)~g_2(\theta),
    \label{decompose equation}
\end{equation}

where 
\begin{equation}
h^*_1(\phi)=\frac{1}{2\pi}, ~ 0\leq \phi<2\pi,
    \label{uni dist}
\end{equation}

and

\begin{equation}
h^*_2(\theta)=\frac{1}{2\pi}\left[1+\frac{r}{R} \cos{\theta}\right], ~ 0\leq \theta<2\pi. 
    \label{cos dist}
\end{equation}

The cumulative distribution function for $\theta$ is
$$
H^*_2(\theta)=\frac{1}{2\pi}\left[\theta+\frac{r}{R} \sin{\theta}\right], ~ 0\leq \theta<2\pi.
$$
\cite{diaconis2013sampling}  use the acceptance-rejection sampling method for  generating samples from the density $h^*_2(\theta)$, and the  algorithm  for the same is also provided in their article.\\

\newpage
\textbf{Proof of the Theorem-\ref{coditional mean direction thm}}

\begin{proof}
To prove the Theorem-\ref{coditional mean direction thm} we need  the following lemmas
\begin{lemma}
The marginal probability density function of $\Theta$ is given by
$$h_6(\theta)=\frac{1}{2\pi} \big[1+ \nu \cos{(\theta-\mu_1)}\big], $$ $0\leq \theta <2\pi$, $0 < \nu \leq 1$.
\label{marginal of theta of dependent}
\end{lemma}

\begin{proof}
    The marginal probability density function of $\Theta$ can be obtained by
  \begin{eqnarray}
      h_6(\theta) &=&\frac{1}{4\pi^2}   \int_{0}^{2\pi} \big[1+ \nu \cos{(\theta-\mu_1)}\big]\big[1-\kappa \sin(\phi-\mu_2+\lambda (\theta-\mu_1))\big]  \,d\phi \nonumber\\
        &=& \frac{1}{4\pi^2}   \big[1+ \nu \cos{(\theta-\mu_1)}\big]  \big[\theta+\kappa \cos(\phi-\mu_2+\lambda (\theta-\mu_1)) \big]_{0}^{2\pi} \nonumber\\
          &=& \frac{1}{2\pi}   \big[1+ \nu \cos{(\theta-\mu_1)}\big] 
        \label{thet1 marginal cal}
  \end{eqnarray}
  Hence, the lemma.
\end{proof}

\begin{lemma}
The conditional probability density function of $\Phi$ given $\Theta=\theta$ is
$$h_5(\phi|\theta)=\frac{1}{2\pi}\big[1-\kappa \sin((\phi-\mu_2)+\lambda(\theta-\mu_1))\big], $$ where $0\leq \phi,\theta <2\pi$,  $0\leq \mu_1,\mu_2 <2\pi$, $-1 \leq \kappa \leq 1$, $\lambda \in \mathbb{R}.$
\label{marginal of theta}
\end{lemma}

\begin{proof}
The conditional probability density function of $\Phi$ given $\Theta=\theta$ can be found using the  joint probability density function from Equation-\ref{dependent pdf} and marginal probability density function of $\theta$ from Lemma-\ref{marginal of theta} which is given by 

  \begin{eqnarray}
      h_5(\phi|\theta) &=&\frac{ \frac{1}{4\pi^2}  \big[1+ \nu \cos{(\theta-\mu_1)}\big]\big[1-\kappa \sin(\phi-\mu_2+\lambda (\theta-\mu_1))\big] }{\frac{1}{2\pi}   \big[1+ \nu \cos{(\theta-\mu_1)}\big] }    \nonumber\\
        &=& \frac{1}{2\pi}\big[1-\kappa \sin((\phi-\mu_2)+\lambda(\theta-\mu_1))\big]
        \label{phi conditional cal}
  \end{eqnarray}
  This completes the lemma.
\end{proof}

    The Equation-\ref{phi conditional cal} can be written as
\begin{eqnarray}
     h_5(\phi|\theta) &=&\frac{1}{2\pi}\left[1-\kappa \sin((\phi-\mu_2)+\lambda(\theta-\mu_1))\right]\nonumber\\
     &=&\frac{1}{2\pi}\left[1+\kappa \cos \left(\frac{3\pi}{2}-\left[(\phi-\mu_2)+\lambda(\theta-\mu_1)\right] \right)\right]\nonumber\\
     &=&\frac{1}{2\pi}\left[1+\kappa \cos \left(\phi-\left(\frac{3\pi}{2}+\mu_2-\lambda(\theta-\mu_1) \right)\right)\right],
     \label{conditional for phi modified}
\end{eqnarray}
which represents Cardioid distribution  (see  \cite{jammalamadaka2001topics}, ) with concentration parameter $-1\leq \kappa \leq 1$, and mean direction $\mu_{\phi|\theta}= \left[\frac{3\pi}{2}+\mu_2+\lambda(\theta-\mu_1)\right] \mod 2\pi $. This proves the theorem.
\end{proof}

\textbf{Proof of the Theorem-\ref{conditional of theta thm}}
\begin{proof}
  Without loss of generality, let us consider $\mu_1=\mu_2=0$. The marginal probability density function of $\Phi$ is given by
  \begin{eqnarray}
      h_5(\phi) &=&\frac{1}{4\pi^2}   \int_{0}^{2\pi} \big[1+ \nu \cos{(\theta)}\big]\big[1-\kappa \sin(\phi+\lambda \theta)\big]  \,d\theta  
 \nonumber\\
        &=& \int_{0}^{2\pi} \left(-\frac{{\kappa}{\nu} \cos\left({\theta}\right) \sin\left({\lambda}{\theta} + {\phi}\right)}{4\pi^{2}} - \frac{{\kappa} \sin\left({\lambda}{\theta} + {\phi}\right)}{4\pi^{2}} + \frac{{1+\nu} \cos\left({\theta}\right)}{4\pi^{2}} \right) \mathrm{d}{\theta} . \nonumber\\
        &=& \frac{1}{2\pi}-  I_1-I_2,
        \label{phi marginal cal1}
  \end{eqnarray}
  where $I_1=\displaystyle \int_{0}^{2\pi}\frac{{\kappa}{\nu} \cos\left({\theta}\right) \sin\left({\lambda}{\theta} + {\phi}\right)}{4\pi^{2}} ~~d\theta$, and 
  $I_2= \displaystyle \int_{0}^{2\pi}\frac{{\kappa} \sin\left({\lambda}{\theta} + {\phi}\right)}{4\pi^{2}} ~~d\theta.$

Now, consider $I_1$
\begin{eqnarray}
      I_1 &=&\displaystyle \int_{0}^{2\pi}\frac{ \cos\left({\theta}\right) \sin\left({\lambda}{\theta} + {\phi}\right)}{4\pi^{2}} ~~d\theta \nonumber\\
      &=&\frac{{\kappa}{\nu}}{8\pi^2} \int_{0}^{2\pi} \left[\sin\left(\left({\lambda} + 1\right) {\theta} + {\phi}\right) + \sin\left(\left({\lambda} - 1\right) {\theta} + {\phi}\right)\right] \, \mathrm{d}{\theta}\nonumber\\
      &=& - \left[\frac{\cos\left(\left({\lambda} + 1\right) {\theta} + {\phi}\right)}{2 \left({\lambda} + 1\right)} +\frac{\cos\left(\left({\lambda} - 1\right) {\theta} + {\phi}\right)}{2 \left({\lambda} - 1\right)}\right]_{0}^{2\pi}
        \label{phi marginal cal I1}
  \end{eqnarray}

Now, consider $I_2$
\begin{eqnarray}
      I_2&=&\displaystyle \int_{0}^{2\pi}\frac{ \cos\left({\theta}\right) \sin\left({\lambda}{\theta} + {\phi}\right)}{4\pi^{2}} ~~d\theta \nonumber\\
      &=&-\frac{{\kappa}}{4\pi^2} \left[\frac{\cos\left({\lambda}{\theta} + {\phi}\right)}{\lambda} \right]_{0}^{2\pi}
        \label{phi marginal cal I2}
  \end{eqnarray}
Using Equation-\ref{phi marginal cal I1} and \ref{phi marginal cal I2} in Equation-\ref{phi marginal cal1} and simplifying we get

\begin{eqnarray}
      h_5(\phi) &=&\frac{1}{2\pi}-\frac{\left({\kappa}{\lambda}^{2} {\nu} + {\kappa}{\lambda}^{2} - {\kappa}\right) \cos\left({\phi} + 2\pi{\lambda}\right) + \left(-{\kappa}{\lambda}^{2} {\nu} - {\kappa}{\lambda}^{2} + {\kappa}\right) \cos\left({\phi}\right)}{4\pi^{2} {\lambda} \left({\lambda}^{2} - 1\right)}\nonumber\\
      &=&  \frac{1}{2\pi}- \left[\frac{\kappa \{\lambda^2(1+\nu)-1\}}{2\pi^2(\lambda^3-\lambda)} \sin{(\lambda\pi)} \right]\sin(\phi+\lambda\pi)
               \nonumber\\
      &=&\frac{1}{2\pi} \left[1 - A \sin(\phi+\lambda\pi)\right]\nonumber\\
      &=&\frac{1}{2\pi}\left[1+A \cos \left(\phi-\left(\frac{3\pi}{2}-\lambda \pi \right)\right) \right],
        \label{phi marginal cal2}
  \end{eqnarray}
where $A=\left[\frac{\kappa \{\lambda^2(1+\nu)-1\}}{\pi(\lambda^3-\lambda)} \sin{(\lambda\pi)} \right]$, $0\leq \phi <2\pi$, $0 < \nu \leq 1$, $-1 \leq \kappa \leq 1$, $\lambda \in \mathbb{R},$ here $\left(\frac{3\pi}{2}-\lambda \pi \right) \mod 2\pi$ is the mean direction.
From the joint density in Equation-\ref{dependent pdf} and the marginal density in Equation-\ref{phi marginal cal2} we obtain the desired conditional density $ \Theta $ given $\Phi=\phi$ as  
$$ h_6(\theta|\phi)=\frac{ \frac{1}{4\pi^2}  \big[1+ \nu \cos{(\theta-\mu_1)}\big]\big[1-\kappa \sin(\phi-\mu_2+\lambda (\theta-\mu_1))\big] }{  \frac{1}{2\pi}\left[1+A \cos \left(\phi-\mu_3\right) \right]}$$
where $0\leq \phi,\theta <2\pi$,  $0 < \nu \leq 1,$ $\mu_3=\left(\frac{3\pi}{2}-\lambda \pi \right)$ $ \mod 2\pi$, $0\leq \mu_1,\mu_2,\mu_3 <2\pi$, $-1 \leq \kappa \leq 1$, $\lambda \in \mathbb{R}.$

\end{proof}

\textbf{Proof of the Theorem-\ref{EAU_thm}}

\begin{proof}
    
  \textbf{Part-I:}
    In this case, we consider $U<p(\Theta)$, and for $\Theta>\pi$ or $\Theta<\pi$ we have $Y=\Theta.$ Hence, we have

    \begin{eqnarray}
        P(Y \leq y)&=& \int_{0}^{y} \frac{1}{2\pi} P(U<p(\theta))  \,d\theta \nonumber\\
        &=&\frac{1}{2\pi} \int_{0}^{y} (1+\nu\cos \theta) \,d\theta .
        \nonumber
    \end{eqnarray}
 Therefore, the integral gives
$
P(Y \leq y)= \frac{1}{4\pi} \left(y+\nu \sin{y}\right)
$

 \textbf{Part-II:} 
    In this case, we consider $U>p(\theta)$. Hence, we have
    \begin{equation*}
Y = \left\{
        \begin{array}{ll}
            \pi-\Theta & \text{when} \quad \Theta<\pi\\
            3\pi-\Theta & \text{when} \quad \Theta>\pi
        \end{array}
    \right.
\end{equation*}
Now, when $0<\Theta<\pi$ then $0<Y<\pi$ and we have 

\begin{eqnarray}
    P(Y \leq y)&=& \int_{\pi-y}^{\pi} \frac{1}{2\pi} P(U>p(\theta))  \,d\theta \nonumber\\
    &=&\frac{1}{4\pi} \int_{\pi-y}^{\pi} (1-\nu\cos \theta) \,d\theta.
    \nonumber
\end{eqnarray}
Therefore, the integral gives
$
P(Y \leq y)= \frac{1}{4\pi} \left(y+\nu \sin{y}\right)
.$
Again, when $\pi<\Theta<2\pi$ then $\pi<Y<2\pi$,we get

\begin{eqnarray}
    P(Y \leq y)&=& \int_{0}^{\pi} \frac {1}{2\pi} P(U>p(\theta)) \,d\theta +\int_{3\pi-y}^{2\pi} \frac{1}{2\pi} P(U>p(\theta))  \,d\theta\nonumber\\
   & =&\frac{1}{4\pi} \int_{0}^{\pi} (1-\nu\cos \theta) \,d\theta+ \frac{1}{4\pi} \int_{3\pi-y}^{2\pi} (1-\nu\cos \theta) \,d\theta
    \nonumber
\end{eqnarray}
Therefore, the integral gives
$
P(Y \leq y)= \frac{1}{4\pi} \left(y+\nu \sin{y}\right)
.$
Adding the two probabilities in \textbf{Part-I} and \textbf{Part-II} we get
$$
H_2^*(y)=P(Y \leq y)= \frac{1}{2\pi} \left(y+\nu \sin{y}\right)
.$$
which has the density function function of Cardioid distribution as in Equation-\ref{Cardioid_pdf} $$h_2^*(y)=\frac{1}{2\pi}(1+\nu\cos y)$$
\end{proof}

\newpage
\textbf{Pseudocode for Theorem-\ref{EAU_thm}}
 \SetKwComment{Comment}{/* }{ */}
\RestyleAlgo{ruled}
\begin{algorithm}
\caption{Pseudocode for the proposed algorithm}\label{algo_eau}
\label{sec_algo}
\KwData{  $X_i \sim  \mbox{~Uniform~}[0,2\pi]$ for $i=1, \cdots, n$.}

$a \in (0,1)$\;
$p_x= \dfrac{1+a \cos{\mathbf{X}}}{2} $\Comment*[r]{define the probability}
$ \mathbf{Y} = \mathbf{0}$ \Comment*[r]{n component vector}
\For{$i = 1 \mbox{~ to~} n$}{
$R_p[i] \sim Bernoulli(p_x[i])$\;
  \eIf{$X[i]<\pi$ }{
    $Z_1[i]=  (X[i]*R_p[i])+(\pi-X[i])*(1-R_p[i])$\;
  }{\If{$X[i]>\pi$}{
      $Z_2[i]=  (X[i]*R_p[i])+(3\pi-X[i])*(1-R_p[i])$\;
    }
  }
  $Y[i] = Z_1[i]+Z_2[i] $\Comment*[r]{$Y \sim F(y)=\dfrac{\left(y+a \sin{y}\right)}{2\pi}.$ }
}
\KwResult{ $Y$  follows Cardioid distribution.}
\end{algorithm}

%% For citations use: 
%%       \cite{<label>} ==> [1]

%%

%% If you have bib database file and want bibtex to generate the
%% bibitems, please use
%%

%%%%%%%%%%%%%%%%%%%%%%%%%%%%%%%%%%

% \newpage
% \begin{scriptsize}
% \bibliographystyle{natbib}
% 	%\bibliographystyle{abbrvnat}
% 	\bibliography{buddha_bib.bib}
% \end{scriptsize}

\end{document}